\ifx\fiverm\undefined
        \newfont\fiverm{cmr5}
\fi
\input prepictex
\input pictex
\input postpictex
\catcode`\@=11
\@input{picmore.tex}
\documentstyle[12pt]{article}

\newcommand{\beq}{\begin{equation}}
\newcommand{\eeq}{\end{equation}}
\newcommand{\beqa}{\begin{eqnarray}}
\newcommand{\eeqa}{\end{eqnarray}}
\newcommand{\nn}{\nonumber}
\newcommand{\eq}[1]{(\ref{#1})}

\newcommand{\define}{\stackrel{\mbox{def.}}{=}}
\newcommand{\ft}{\stackrel{\mbox{F.T.}}{=}}
\newfont{\sgcal}{eufm9}
\newfont{\smcal}{eusm9}
\newfont{\smbf}{msbm9}
\newfont{\gcal}{eufm10}
\newfont{\mcal}{eusm10}
\newfont{\mbf}{msbm10}
\newfont{\Gcal}{eufm10 scaled\magstep1}
\newfont{\Mcal}{eusm10 scaled\magstep1}
\newfont{\Mbf}{msbm10 scaled\magstep1}
\newcommand{\ra}{\rightarrow}
\newcommand{\lra}{\leftrightarrow}
\newcommand{\der}{\partial}
\newcommand{\eps}{\varepsilon}

\newcommand{\I}{\mbox{\Mbf I}}
\newcommand{\J}{\mbox{\Mbf J}}
\newcommand{\V}{\mbox{\Mbf V}}
\newcommand{\Omat}{\mbox{\Mbf O}}
\newcommand{\K}{\mbox{\Mcal K}}
\newcommand{\F}{\mbox{\Mcal F}}
\newcommand{\matF}{\mbox{\Mbf F}}
\begin{document}
\begin{titlepage}
\setcounter{page}{0}
\renewcommand{\thefootnote}{\fnsymbol{footnote}}


\begin{flushright}
\mbox{PUPT-2018}\\
\mbox{KAIST-TH 2001/17}\\
\mbox{hep-th/0112176}\\
\end{flushright}

\vspace{0mm}
\begin{center}
{\Large \bf One-loop Noncommutative 
$U(1)$ Gauge Theory from Bosonic Worldline Approach} 
\vspace{5mm}

{\large Youngjai Kiem$^{a,c}$\footnote{Current address: 
Department of Physics, Korea Advanced Institute 
of Science and Technology, Taejon 305-701, Korea; E-mail: ykiem@muon.kaist.ac.kr}, 
Yeonjung Kim$^{b}$,
        Cheol Ryou$^{a}$, {\sc and} Haru-Tada Sato$^{a}$ } \\
\vspace{10mm}
{\em $^a$ BK21 Physics Research Division and Institute of 
Basic Science, \\ 
Sungkyunkwan University, Suwon 440-746, Korea} \\
\vspace{4mm}
{\em $^b$ Department of Physics, Korea Advanced Institute 
of Science and Technology, \\ 
Taejon 305-701, Korea} \\
\vspace{4mm}
{\em $^c$ Physics Department, Princeton University, 
Princeton, NJ08544, USA} \\
\vspace{4mm}
\centerline{\tt ykiem, cheol, haru@newton.skku.ac.kr, 
             \hskip0.3cm 
             geni@muon.kaist.ac.kr}
\end{center}

\vspace{4mm}

\begin{abstract}
We develop a method to compute the one-loop effective
action of noncommutative $U(1)$ gauge theory based on
the bosonic worldline formalism, and derive
compact expressions for $N$-point 1PI amplitudes.  
The method, resembling perturbative string computations, 
shows that open Wilson lines emerge as a gauge invariant
completion of certain terms in the effective action.
The terms involving open Wilson lines are of the form 
reminiscent of closed string exchanges between the 
states living on the two boundaries of a cylinder.  
They are also consistent with recent matrix theory  
analysis and the results from noncommutative scalar 
field theories with cubic interactions.
\end{abstract}

\vspace{3mm}
\vfill
\begin{flushleft}
PACS: 11.15.Bt, 11.25.Mj, 11.10.-z\\
Keywords: noncommutative gauge theory, open Wilson lines,
Worldline formalism
\end{flushleft}

\end{titlepage}
\newpage
\setcounter{footnote}{0}
\renewcommand{\thefootnote}{\arabic{footnote}}
\renewcommand{\theequation}{\thesection.\arabic{equation}}

\section{Introduction}\label{sec1}
\setcounter{section}{1}
\setcounter{equation}{0}
\indent

Noncommutative field theories are a miniaturized version
of string theory, through which we can discuss 
issues such as nonlocality and off-shell physics 
in a controlled fashion (for reviews, 
see \cite{reviews}).  A notable aspect in this regard 
is the appearance of open Wilson lines \cite{owl}; 
they allow us to form off-shell gauge invariant 
observables \cite{trek2}, and capture the dipole 
nature of noncommutative field theories representing 
their inherent nonlocality \cite{dipole}.  Typically, 
the low energy effective description of noncommutative
D-branes corresponds to noncommutative gauge theory,
a prime example of noncommutative field theories.  
Even in its simplest setup of noncommutative $U(1)$
gauge theory, the complete computation of the one-loop 
effective action is non-trivial.  In particular,
to directly test if open Wilson lines emerge as 
expected, one has to sum over an infinite number of 
gauge field insertions.  An efficient method that
produces the manageable form of the $N$-point 1PI
amplitudes and further allows the summation over
$N$ is thus desirable.  The development of such 
a method for noncommutative $U(1)$ gauge theory 
at one-loop is the task achieved in this paper 
following the bosonic worldline 
approach \cite{bernkoso}-\cite{HTS}.  We derive 
a scheme that resembles perturbative string  
computations of amplitudes involving gauge boson 
vertex operators.  The main difference is that 
being a field theory construction,
our method is valid off-shell, as well as on-shell.
The worldline formalism, that
has been known to produce string-theory-like schemes 
in the commutative context \cite{St}, turns out 
to be a useful device that allows us to keep track 
of all possible 
terms in the effective action.  It is especially helpful
when taking care of various contact terms that 
appear in the case of noncommutative $U(1)$ gauge
theory, whose appearance is rather similar to the 
case of nonabelian gauge theories in commutative 
space-time.   

The $N$-point terms in the one-loop effective action
computed by our method are summarized by 
Eqs.~(\ref{Wick})-({\ref{propt})
for the ghost loop contributions, and by
Eqs.~(\ref{pinch}), (\ref{Wick2}) and (\ref{gluon})
for the gauge loop contributions.  They can be
compactly rewritten as (\ref{Wick3}) and 
(\ref{gwick}).
After summing over certain class of (an infinite number
of) terms in the effective action, we find that
it contains the sequence of terms: 
\begin{eqnarray}
 \Gamma & = & 
  \frac{D-2}{2} \int \frac{ d^D k }{ (2 \pi)^D }
   W_{k} [A]  {\cal K}_0 (k) W_{-k} [A] \nonumber \\ 
  &  & +
\frac{1}{2} \sum_{n_1 ,  n_2 }  
   \frac{(-1)^{n_1}}
 {n_1 ! n_2 !} \int \frac{d^D k}{(2 \pi )^D}
  \F_{k}^{(n_1)} [A] {\cal K}_{n_1 + n_2} (k) 
  \F_{-k}^{(n_2)} [ A ] ~ ,
\label{u1eff}
\end{eqnarray}
where the $n$-th descendent $\F_k^{(n)}$ 
( $ W_k [A ] = \F_k^{(0)}$) of 
open Wilson lines 
involving $n$-copies of field strength tensor
$F_{\mu \nu}  = \partial_{\mu} A_{\nu} -  
   \partial_{\nu} A_{\mu} -ig ( A_{\mu} \star
 A_{\nu} - A_{\nu} \star A_{\mu} )$ is
defined as 
\begin{equation}
 \F_k^{(n)} = 2^n  (ig)^n \int d^D x 
  P_{\star} \left[ W(x, C ) \prod_{i=1}^n F (x) \right] 
 \star e^{ i k x}  
\end{equation}
in terms of the `straight' open Wilson line \cite{owl}
\begin{equation}
 W( x, C) = P_{\star} \exp 
 \left ( i g \int_0^1 dt \frac{\partial y^{\mu} (t) }
  {\partial t}
   A_{\mu} ( x + y(t) ) \right) ~~~ , ~~~
 y^{\mu} (t) = \theta^{\mu \nu} k_{\nu} t ~ .
\label{vwilson}
\end{equation}
Basic notations are given at the end of this
section, the space-time indices are explained
in detail later in this paper, and 
$P_{\star}$ is the path-ordering with respect to
the $\star$-product.  The explicit expression for
the kernel ${\cal K}_n$ is given in (\ref{prop}).
The summation $\sum_{n_1 , n_2}$ represents
$\sum_{n_1 = 0}^{\infty} \sum_{n_2 = 0}^{\infty}$
excluding the case $n_1 = n_2 = 0$; the contribution
coming from $n_1 = n_2 =0$ is separated out into
the first term of (\ref{u1eff}), and it also 
contains the ghost loop 
contribution\footnote{In noncommutative real scalar
field theory with a cubic 
interaction \cite{owl1,ana1,ana2,owl2}, 
the whole one-loop
effective action in the large noncommutativity
limit can be summed up to the form of
$n_1 = n_2 = 0$ term with coefficient $1/2$.  
In this context, the open 
Wilson line is replaced with a scalar analog of 
the vector Wilson line (\ref{vwilson}).  
The kernel ${\cal K}_0$ in this case is the 
leading term in the asymptotic expansion of
${\cal K}_0$ in (\ref{u1eff}) for the large value 
of the argument.}.  
Each term in
(\ref{u1eff}) involves an infinite number of 
higher-point terms originating from the expansion of
the exponential function $W(x,C)$ 
and extra higher-point 
terms from the noncommutative
commutator term in the field strength tensor.
Partially due to the factor 
$(-1)^{n_1}$, all the terms in (\ref{u1eff}) 
vanish in the commutative limit, which is not
the case in the noncommutative setup. 
When there are more fields transforming 
adjointly under the noncommutative gauge
group, it was argued in \cite{mvr,armoni} that
the coefficient $(D-2)$ in (\ref{u1eff}) is 
in general replaced
with $(N_B - N_F)$ where $N_B$ ($N_F$) is the 
bosonic (fermionic) degrees of freedom
(see also \cite{Minwalla,ncft}).  
In the maximally supersymmetric ${\cal N} = 4$, 
the terms up to $n_1 + n_2 < 4$ vanish, as the 
consequence of boson-fermion 
cancellations \cite{kl}.  

Our results (\ref{u1eff}) are consistent
with the gauge invariant completion suggested
in Ref.~\cite{trek2}.  
In \cite{trek1}, the leading nonvanishing
four-point terms in the $D=4$, ${\cal N} = 4$ 
noncommutative supersymmetric Yang-Mills theory were
computed by adopting the perturbative string
theory technique and taking the Seiberg-Witten
limit \cite{sw}.  They turned out to be the type of 
the $n_1 + n_2 = 4$ terms of (\ref{u1eff})
when taking $W (x, C) =1$ and neglecting the 
commutator terms, modulo the detailed space-time
index structure on which we will comment later.  
Based on the requirement
of (off-shell) gauge invariance, it was conjectured
in \cite{trek2} that the higher point terms that
make up the $\F_k^{(n)}$ should be present 
in the one-loop effective action.  Further 
perturbative evidence supporting this conjecture
was provided by Refs.~\cite{trek3}.  
More recently, the matrix theory side considerations
of open Wilson lines in noncommutative gauge
theory were reported in \cite{mvr} for one-loop case
and in \cite{texas} for two-loop case.  
In addition, the authors of \cite{armoni} 
obtained $n_1 = n_2 =0$ 
result of (\ref{u1eff}) by explicitly summing up 
certain Feynman diagrams.  It should be noted 
however that
there are extra terms in the effective action that can
not be easily represented in terms of open Wilson
lines, as our analysis will show.  
We will make further comments on them later.
It is an interesting outstanding problem to see
if they can also be written in terms of (variants of)
open Wilson lines, and to see if they are
absent in the supersymmetric gauge theories.
   
The analysis of $U(1)$ gauge theory closely 
parallels the analysis of scalar field theory
with a cubic interaction reported in 
\cite{owl1,ana1,ana2,owl2}, where
it was pushed to two-loops.  As is clear 
from the comparison of the final answers, our
results indicate that the appearance of open Wilson 
lines is a universal feature of noncommutative field 
theories regardless of detailed spin 
contents \cite{klp}.
Main physical features found 
in \cite{owl1,ana1,ana2,owl2}
are still present without any essential modifications;
for example, the terms (\ref{u1eff}) are of the
form reminiscent of `closed string' exchanges
between the states living on the two boundaries
of a cylinder.
There is, however, an important difference.  
In the case of the scalar noncommutative
field theories, to obtain the scalar version of
open Wilson lines, one needs to take a large
noncommutativity limit to reduce the `closed
string kernel' ${\cal K}_{n_1 + n_2}$ to its leading
term in the asymptotic expansion.  In the 
gauge theory case, it is not necessary to
take a similar limit; the summation over 
$N$-point functions goes through for finite
value of the noncommutativity parameter.
It suggests that the appearance of open
Wilson lines is closely related to the existence
of $U(\infty )$ symmetry of a theory in
consideration.  

This paper is organized as follows.  In section 2,
we review the action of noncommutative $U(1)$
gauge theory in the context of the background field
method \cite{background}.  
The choice of the background field gauge 
allows us to formulate a scheme with manifest
gauge invariance that closely
resembles the perturbative string theory
computations.  
In section 3, we evaluate the $N$-point
functions resulting from the ghost loop and the
gauge loop, adopting the bosonic worldline formalism.
Particular attentions are paid to the treatment of
various contact terms that are essential in maintaining
the gauge invariance.  Some of the details along
this line are presented in Appendix B and C.  
A detailed implementation of noncommutative worldline
formalism is relegated to Appendix A.  
Eventually, we find an expression for the
ghost loop that involves
gauge boson vertex operators similar to the
vertex operators of bosonic string theory.  
For the gauge loop, appropriate
vertex operators turn out to be similar to the
0-picture gauge boson vertex operators in superstring
theory.  In section 4, we sum certain class of $N$-point 
terms to obtain (\ref{u1eff}), showing the emergence of
open Wilson lines and their descendents in the
one-loop effective action.  

The notations adopted throughout 
this paper are as follows.  We consider the 
noncommutative $U(1)$
gauge theory on a noncommutative plane $R^D$, where 
the coordinates satisfy 
\beq
[\,x^\mu \,,x^\nu\,] = i\theta^{\mu\nu} \label{com} ~ .
\eeq
Via the Weyl-Moyal correspondence, the product between
fields is given by the $\star$-product 
\beq
 \phi \star \phi(x) = 
e^{{i\over2}\theta^{\mu\nu}\der^y_\mu\der^z_\nu}\,
\phi(y)\,\phi(z)\,\Bigr|_{y=z=x} \ .  \label{star}
\eeq
In addition, the following notations are used:
\beq
p \wedge k = p_\mu \theta^{\mu\nu} k_{\nu} \ ,\qquad
p \circ k = p_\mu (-\theta^2)^{\mu\nu} k_{\nu} \ge 0 
\ . \label{product}
\eeq
We only consider the space-space noncommutativity in
this paper.  The Lorentz indices, therefore, will be 
frequently put upside down for convenience.

\section{Review: Background field action of
noncommutative $U(1)$ gauge theory}\label{sec2}
\setcounter{section}{2}
\setcounter{equation}{0}
\indent

Our main interest is the computation of one-loop effective
action maintaining the explicit gauge invariance.  
For this purpose, we employ the background 
field method, splitting the gauge field $A$ 
into $A+Q$, where $A$ and $Q$ are the background fields
and quantum fluctuations, respectively. 
The noncommutative $U(1)$ gauge theory is then 
described by the following action: 
\beq
S = \int d^Dx \Bigl\{\,
-{1\over 4}F_{\mu\nu}(A+Q)\star F^{\mu\nu}(A+Q) 
-{1\over {2\alpha}}(\bar D^{\mu}Q_{\mu})^2_\star 
+{\bar C}{\bar D}_\mu \star D^{A+Q}_\mu C \,\Bigr\}\ , 
\eeq
where we include ghost field $C$ and anti-ghost field
$\bar{C}$.  Eventually, we will choose the Feynman gauge
setting $\alpha = 1$ (keeping the gauge invariance
for $A$).  
The covariant derivatives and field strength for noncommutative 
gauge fields $X_\mu$ (either $A_\mu$ or $Q_\mu$) are given by 
\beqa
D^X_{\mu}Y &=& \partial_{\mu}Y -ig[X_{\mu}, Y]_{\star}\ , 
\qquad \bar D_\mu \equiv D^A_\mu\ ,\\
{}~F_{\mu\nu}(X)&=&\partial_{\mu}X_{\nu} - \partial_{\nu}X_{\mu}
-ig[X_{\mu},X_{\nu}]_{\star}\ , 
 \qquad \bar{F}_{\mu\nu}\equiv F_{\mu\nu}(A)\ ,
\eeqa
where the $\star$-commutator represents 
$[\,X,Y\,]_\star = X \star Y - Y\star X$.   
It is useful to use the following relation:
\beq
F_{\mu\nu}(A+Q) = \bar{F}_{\mu\nu}+
 \bar{D}_{\mu}Q_{\nu}-\bar{D}_{\nu}Q_{\mu}
-ig[Q_{\mu}, Q_{\nu}]_{\star} \ .
\eeq
After some algebra, we organize the action up to surface terms
\beqa
S &=& \int d^Dx \Bigl[-{1\over 4}\bar{F}^{\mu\nu}
 \star \bar{F}_{\mu\nu} 
- \bar{F}^{\mu\nu}\star \bar{D}_{\mu}Q_{\nu} \nn 
\\
&&-{1\over2}\,(\bar{D}^{\mu}Q^{\nu})\star
  (\bar{D}_{\mu}Q_{\nu})
+ {1\over2}(1-{1\over\alpha})\,(\bar{D}^{\mu}Q^{\mu})
  \star(\bar{D}_{\nu}Q_{\nu}) 
+ig\bar{F}^{\mu\nu}\star[Q_{\mu}, Q_{\nu}]_{\star}\nn 
\\
&&+{1\over4} g^2 [Q^\mu, Q^\nu]_\star \star 
  [Q_\mu, Q_\nu]_\star
+ ig[Q^\mu, Q^\nu]_\star \star \bar{D}_\mu Q_\nu \Bigr] \nn
\\
&&+\int d^Dx\, \Bigl[\, \bar{C}\bar{D}_\mu \bar{D}^\mu C
-ig\bar{C}\bar{D}_\mu [Q^\mu,C]_\star \,\Bigr]_\star \ .
\eeqa
Discarding one-particle-reducible terms and 
setting ${\alpha =1}$, the 
one-loop relevant parts of $S$, needed when computing
the one-loop effective action, read 
$S\sim S^{gauge}+S^{ghost}$ where
\beqa
S^{gauge}  &=& \int d^Dx
\frac{1}{2}Q^{\mu}\star[g_{\mu\nu}\bar{D}^{\rho}
 \star \bar{D}_{\rho}
-4ig\bar{F}_{\mu\nu}] \star Q^{\nu} \ , \label{Sgauge}\\
S^{ghost} &=& \int d^Dx\,
\bar{C} \star\bar{D}_\mu\star \bar{D}^\mu\star C \ .
\eeqa
To effectively deal with the space-time 
nonlocality, it is convenient to adopt the 
momentum basis via Fourier transformation. 
Utilizing the following relations for the terms in 
$S^{gauge}$ 
\beqa
\int d^Dx \bar{D}^{\rho}{Q^{\mu}}\star 
  \bar{D}_{\rho}{Q_{\mu}}&=&
\int d^Dx \Bigl[-{Q^{\mu}}\partial^{2}{Q_{\mu}}+ 2ig
A_{\rho}\star[\partial^{\rho}Q^{\mu},Q_{\mu}]_\star \nn\\
&&+2g^{2}(A^{\rho}B_{\rho}Q^{\mu}Q_{\mu}-A^{\rho}Q^{\mu}A_{\rho}
Q_{\mu})_\star \Bigr]\ , \label{sa1} \\
-2ig\int d^Dx Q^{\mu} \star \bar{F}_{\mu\nu} \star Q^{\nu}
&=& -ig\int d^Dx Q^{\mu} 
 \star [\bar{F}_{\mu\nu}, Q^{\nu}]_\star \nn\\
&=&\int d^Dx Q^{\mu}\star 
 [\bar D_{\mu},\bar D_{\nu}]\star Q^{\nu}\ ,
\label{sa2}
\eeqa
valid up to surface terms, 
we realize that \eq{sa1} produces $S_2$ part, 
and \eq{sa2} yields $S_3$ and $S_4$ parts given by: 
\beq
S^{gauge} = S_2+S_3+S_4 \ , 
\eeq
where 
\beqa
S_2(Q) &=& -{1\over2}\int{d^Dk_1\over(2\pi)^D}
  {d^Dk_2\over(2\pi)^D}
\Bigl\{ik_1^{\mu}-ig\int{d^Dp\over(2\pi)^D}\bigl(e^{{i\over2} 
k_1\wedge p}-e^{-{i\over2} k_1\wedge p}\bigr)
    \hat A^{\mu}(p) \Bigr\}\nn\\
&\times & \Bigl\{ik_{2\mu}-
  ig\int{d^Dp\over(2\pi)^D}\bigl(e^{{i\over2}
k_2\wedge p}-e^{-{i\over2} k_2\wedge p}\bigr)
  \hat A_{\mu}(p) \Bigr\}
\hat Q_{\nu}(k_1)\hat Q^{\nu}(k_2)\ ,\label{S2} \\
S_3(Q) &=& \int{d^Dk_1\over(2\pi)^D}{d^Dk_2\over(2\pi)^D}
\,g\int{d^Dp\over(2\pi)^D}\bigl(e^{{i\over2}
k_1\wedge p}-e^{-{i\over2} k_1\wedge p}\bigr) \nn\\
&\times& 
 \bigl(p_{\mu}\hat A_{\nu}(p)-p_{\nu}\hat A_{\mu}(p)\bigr)
\hat Q^{\mu}(k_1)\hat Q^{\nu}(k_2)\ ,\label{S3} \\
S_4(Q) &=& \int{d^Dk_1\over(2\pi)^D}{d^Dk_2\over(2\pi)^D}
\,g^2\int{d^Dp_1\over(2\pi)^D}{d^Dp_2\over(2\pi)^D} 
\bigl(e^{{i\over2}k_1\wedge p_1}
   -e^{-{i\over2} k_1\wedge p_1}\bigr)
\bigl(e^{{i\over2}k_2\wedge p_2}
   -e^{-{i\over2} k_2\wedge p_2}\bigr)\nn\\
&\times&\bigl(\hat A_{\mu}(p_1)\hat A_{\nu}(p_2)
    -\hat A_{\nu}(p_1)\hat
A_{\mu}(p_2)\bigr) \hat Q^{\mu}(k_1)
  \hat Q^{\nu}(k_2) \ .\label{S4}
\eeqa
In deriving these expressions, we have assumed 
the momentum conservation as usual, and 
$\hat X$ stands for the {}~Fourier 
transforms of $X=Q_\nu,A_\nu$.  
The ghost action $S^{ghost}$ is simply given by 
replacing the $Q$ field 
dependence in $S_2$ with the (anti-) ghost field 
dependence, i.e., symbolically, $S^{ghost}=S_2(C)$. 

When we set
$\theta^{\mu \nu} = 0$, the interaction terms
between $A$ and $Q$ disappear and the theory
becomes trivial, as the commutative $U(1)$ gauge
theory should be.  In the presence of nonvanishing
$\theta^{\mu \nu}$, however, the $U(1)$ gauge theory 
in consideration closely mimics the behavior of
nonabelian gauge theories in commutative 
space-time.  To summarize, the ghost and gauge 
loop contributions to the one-loop effective action 
$\Gamma[A]$ are given by the logarithm of 
formal one-loop
determinants 
\beqa
\Gamma^{\rm ghost}[A] &=& 
  \ln\mbox{Det}\Bigl[\frac{-\delta^2 S_2(C)}
{\delta{\hat C(k_1)}\delta{\hat{\bar C}(k_2)}} 
   \Bigr] \ , \label{gam1}
\\
\Gamma^{\rm gauge}[A] &=& 
  \ln\mbox{Det}^{-1/2}\Bigl[\frac{-\delta^2}
{\delta{\hat Q(k_1)}\delta{\hat Q(k_2)}}(S_2+S_3+S_4) 
   \Bigr] \ , \label{gam2}
\eeqa
which we will evaluate adopting the bosonic
worldline formalism.

\section{Computations of $N$-point functions in 
the one-loop effective action}\label{sec3}
\setcounter{section}{3}
\setcounter{equation}{0}
\indent

In this section, we show how we can extract the worldline path 
integral expressions for $N$-point proper amplitudes.
The starting point for the computation 
is to consider the ghost contribution 
in detail, where the action consisting of only $S_2$
is simpler.  Particular attention will be paid to the 
treatment of contact terms (quartic terms) in $S_2$
and the necessary point-splitting regularization.
Once we understand the ghost loop contribution,
the treatment of the gauge loop is straightforward.
We will eventually show that $S_3$ and $S_4$ parts
combine to form field strength tensors, where
the (quartic) $S_4$ part provides the $\star$-commutator 
terms.  
\subsection{The ghost loop}\label{sec3.1}
\indent

We first consider the ghost loop~\eq{gam1} following 
the `stripping method' developed in the treatment of
noncommutative scalar field theories~\cite{ana2}.  The
details not presented in that reference are available
in Appendix A\footnote{One should note that the 
computations here closely parallel those using 
`nonstripping method' of \cite{ana1} as well. 
It turns out that both methods produce exactly
the same result at the one-loop level.  For the 
treatment of higher loops, however, the stripping
method of \cite{ana2} is more convenient.}. 
The basic idea is to separate out the overall Filk 
phase factor that is responsible for the 
$\star$-products (or $\star^{\nu}$-products in the
one-loop context) between background fields
(stripping). We then attach an extra phase 
factor depending on the loop momentum when
a `nonplanar crossing' happens.  It 
shows up for an external
insertion along the `inner' boundary of a 
one-loop diagram in the double-line 
notation.  See Figure 1, for example. 

Taking into account of stripping factors from \eq{base1} 
and \eq{base2}, 
we extract the effective vertex $G_1 + G_2$ out of the second 
derivatives of $S_2$:
\beqa
G_1(k)&=& k^2+2gk^{\mu}\int{d^Dp\over(2\pi)^D}
          (1-e^{ik\wedge p})\hat A_{\mu}(p)\ , \label{G1}\\
G_2(k)&=&g^2\int{d^Dp_1\over(2\pi)^D}\int{d^Dp_2\over(2\pi)^D}
(1-e^{ik\wedge p_1})(1-e^{i(k+p_1)\wedge p_2})
\hat A_{\mu}(p_1)\hat A_{\nu}(p_2)g^{\mu\nu} \ .\label{G2}
\eeqa
This leads to a path integral expression 
for \eq{gam1} given by
\beq
\Gamma^{\rm ghost} = -\int{dT\over T}\int{\cal D}x {\cal D}k\,
{\rm P}\,\exp\Bigl[\,-\int\limits_0^T d\tau 
 \Bigl\{ k^2-ik{\dot x}
+ G_1(k) + G_2(k) \Bigr\} \,\Bigr] \ . \label{ghint}
 \eeq
The contribution from $G_1$ comes from the three-point
vertex with a single external line (and two 
internal lines) familiar from
$\phi^3$ scalar theory in the background field 
method \cite{ana2}; two terms, $1$ and $-e^{i k \wedge p}$,
represent the outer (without a crossing) and 
inner (with a crossing) insertions in the double-line
notation, respectively.  The factor $-e^{i k \wedge p}$
is the aforementioned extra phase factor depending on
the internal momentum $k$, which was also 
used in \cite{Minwalla}.  
The extra minus sign for the inner
insertion is a property of gauge theories.  The
interaction vertex in $G_2$ (\ref{G2}) involves
two external fields inserted at the same point.
See Figure 1 and, as such, they are what we call 
contact terms.
These terms will play a crucial role in simplifying
the final result and enforcing the gauge
invariance, as will be shown shortly. 

It is convenient to set $G_2 =0$
first, and the resulting effective action
will be called $\Gamma^{(0) ghost}$.
Expanding the $\hat{A}$-dependent piece in (3.1) in
the exponential part of (\ref{ghint}), we have 
\beqa
\Gamma^{(0)\rm ghost} &=& -\int{dT\over T}\int{\cal D}x{\cal D}k
\exp\Bigl[-\int\limits_0^T d\tau 
  \Bigl(k^2-ik{\dot x} \Bigr) \Bigr] \nn\\
&\times& \sum_{N=0}^\infty \sum_{n=0}^N\Bigl(-2g\Bigr)^N\,
\prod_{l=1}^n\int\limits_0^{\tau_{l+1}}d\tau_l
\int{d^Dp_l \over(2\pi)^D}\, \bigl[
  k\cdot{\hat A}_{\star} (p_l) \bigr] \nn\\
&\times&\prod_{j=1}^{N-n}\int\limits_0^{\tau'_{j+1}}d\tau'_j
\int{d^Dp'_j\over(2\pi)^D}\,
\bigl[-k\cdot\hat A_{\star} (p'_j)\bigr]  
 \exp\Bigl[\,-i\sum_{j=1}^{N-n}p'_j \wedge k(\tau'_j)\,\Bigr]\ ,
\eeqa
where $n$ is the number of outer insertions and
$N-n$ is the number of inner insertions.  
In order to 
generate all possible Feynman diagrams automatically, we 
replace the external background fields $\hat{A}$
with sums of all 
plane wave modes (after then, inserting the
factor ${1\over N!}$ is necessary):
\beqa
&&\prod_{l=1}^n\int\limits_0^{\tau_{l+1}}d\tau_l
\int{d^Dp_l \over(2\pi)^D}\, \bigl[ 
 k\cdot{\hat A}_{\star} (p_l) \bigr]
\prod_{j=1}^{N-n}\int\limits_0^{\tau'_{j+1}}d\tau'_j
\int{d^Dp'_j\over(2\pi)^D}\,
\bigl[-k\cdot{\hat A}_{\star} (p'_j)\bigr]  \nn\\
&&\ra \,{1\over N!}\,
\prod_{l=1}^n\int\limits_0^{\tau_{l+1}}d\tau_l
\prod_{j=1}^{N-n}\int\limits_0^{\tau'_{j+1}}d\tau'_j
(k(\tau_1)\cdot\epsilon_1)
 \cdots (k(\tau_N)\cdot\epsilon_N)(-1)^{N-n}\nn\\
&&\Bigl(\, e^{ip_1x(\tau_1)}\star^\nu
e^{ip_2x(\tau_2)}\star^\nu\cdots\star^\nu e^{ip_Nx(\tau_N)} 
+ \quad(\mbox{all $p_i$ permutations})\, \Bigr)\ ,
\eeqa
where $\epsilon_i$ is the polarization vector.
The stripped overall Filk phase shows itself through
$\star^\nu$ defined in \eq{newstar} as  
\beq
\phi\star^\nu\varphi(x) =
e^{{i\over2}\nu\theta^{\mu\nu}\der^y_\mu\der^z_\nu}\,
\phi(y)\,\varphi(z)\,\Bigr|_{y=z=x} 
\qquad\mbox{for}\quad\nu=0,\pm1\ ,
\eeq
where $\nu = 1$ when both $\phi$ and $\varphi$ are
outer insertions, $\nu = -1$ when both of them are
inner insertions, and $\nu = 0$ otherwise.
The $\star$-products at tree level are precisely
the $\nu=1$ (outer boundary) $\star^{\nu}$-
products.  
The quantities with primes represent inner insertions.
When compared to scalar field theories with
a cubic interaction, there is an 
extra $(-1)^{N-n}$ factor (Cf. \eq{p3rule}). 
To summarize the result so far, the $N$-point 
contributions are given by
\beqa\label{GN0C}\Gamma_N^{(0)\rm ghost}
&=&-\left({-2g}\right)^N \sum_{\{\nu_i\}}
\int{dT\over T}\left(\prod_{i=1}^N\int\limits_0^Td\tau_i\right)
e^{{i\over4}\sum_{i<j}p_i\wedge p_j (\nu_i+\nu_j)\eps(\tau_{ij})}
\left(\prod_{j=1}^N (-)^{\alpha_j}\right)
 \nn \\ 
&&\times \int{\cal D}x e^{i\sum_ip_ix(\tau_i)}
   \int{\cal D}k 
(k(\tau_1)\cdot\epsilon_1)\cdots(k(\tau_N)\cdot\epsilon_N) 
  \nn \\
&& \times \exp \left( 
 -\int\limits_0^T(k^2(\tau)-ik{\dot x}(\tau))d\tau \right)
\prod_{j=1}^Ne^{-i\alpha_j p_j \wedge k(\tau_j)}\ ,
\eeqa
which is to be summed with the weight
$1/N!$ to become $\Gamma^{(0)ghost}$.   
For inner insertions, we have $\alpha_i = 1$, $\nu_i = -1$,
and for outer insertions, $\alpha_j = 0$, $\nu_j = 1 $ 
(defined in \eq{aj}).   The summation
$\sum_{\nu_i}$ is the summation over all possible
inner/outer insertions consisting of $2^N$ terms.    
The summation over all possible 
permutations of external momenta is
encoded in the integration range of $\tau$ variables,
and the $\star^{\nu}$ phase factor effect gives the phase
factor in the first line where $\eps ( \tau_{ij} )
 = {\rm sign} (\tau_i - \tau_j )$. 

The next task is the evaluation of the path integral
over $k(\tau )$.  If we replace the products of 
$k ( \tau_j ) \cdot \epsilon_j $'s with the functional
derivatives $\epsilon^{\mu}_j 
  \delta / \delta (i\dot x_{\mu}(\tau_j))$,
the resulting path integral is a Gaussian type that can 
be straightforwardly evaluated:
\beqa\label{GN0C2}
\Gamma_N^{(0)\rm ghost}&=&-\left({-2g}\right)^N \sum_{\{\nu_i\}}
\int{dT\over T}\left(\prod_{i=1}^N\int\limits_0^Td\tau_i\right)
e^{{i\over4}\sum_{i<j}p_i\wedge p_j 
 (\nu_i+\nu_j)\eps(\tau_{ij})} 
  \left(\prod_{j=1}^N (-)^{\alpha_j}\right) \nn\\   
&\times&
\int{\cal D}x\, e^{i\sum_ip_ix(\tau_i)}  
\prod_{j=1}^N
\epsilon_j^{\mu}\,\frac{\delta}
 {\delta (i\dot x_{\mu}(\tau_j))}\, K\ ,
\eeqa
where $K$ is given by 
\beq\label{K}
K={\cal N}(T)\, e^{-{1\over4}\int\limits_0^T{\dot x}^2 d\tau}
\prod_{j=1}^N\, e^{-{1\over2}{\dot x(\tau_j)}
  \wedge p_j \alpha_j} 
\eeq
with the normalization factor ${\cal N}(T)$:
\beq\label{NT}
{\cal N}(T)=\int{\cal D}k\, 
  e^{-\int\limits_0^T k^2 d\tau}\ , \qquad
{\cal N}(T)\hskip-5pt
  \int\limits_{x(0)=x(T)}\hskip-15pt{\cal D}x\,
e^{-{1\over4}\int\limits_0^T {\dot x}^2 d\tau}
=\left({1\over4\pi T} \right)^{D\over2}\ . 
\eeq
The functional derivatives in (\ref{GN0C2}) acting
on the exponential part of $K$ generate large
number of terms when $N >1$.  In particular, the
derivative operator can hit the $\dot x^{\nu} (\tau_j )$
already taken out from the exponential part 
\begin{equation}
  \frac{\partial}{\partial \dot{x}_\nu 
   (\tau_i)}\dot{x}^\mu(\tau_j)
=g^{\nu\mu}\delta(\tau_i-\tau_j) 
\end{equation}
producing extra `contact terms'.  There are progressively
many terms of this kind as $N$ increases, and we write
down the result after taking the functional derivatives as  
\beqa\label{aftkint}
\Gamma_N^{(0)\rm ghost}&=& -\left({-ig}\right)^N \sum_{\{\nu_i\}}
\int{dT\over T}\left(\prod_{i=1}^N\int\limits_0^Td\tau_i\right)
e^{{i\over4}\sum_{i<j}p_i\wedge p_j (\nu_i+\nu_j)\eps(\tau_{ij})}
\left(\prod_{j=1}^N (-)^{\alpha_j}\right) \nn\\
&\times& {\cal N}(T)
 \int{\cal D}x\, e^{-\int\limits_0^T{1\over4}{\dot x}^2
(\tau)\,d\tau} \prod_{j=1}^N 
  {\cal O}_j(\tau_j) \,+\, \cdots  \ , 
\eeqa
where the gauge particle insertion in a ghost loop 
is given by ${\cal O}_j(\tau)$ defined as
\beq\label{OJ}
{\cal O}_j(\tau)= \epsilon_j^{\rho}
\Bigl(\, \dot x^{\rho}(\tau) \,+\, \Theta^\rho(\tau) \,\Bigr)
\exp\Bigl[\, i p_j^\mu \Bigl(x_\mu(\tau) -
{i\over2}\alpha_j \theta^{\mu\nu}{\dot x}^\nu(\tau) 
  \Bigr) \,\Bigr]\ ,
\eeq
and
\beq\label{defth}
\Theta^\mu(\tau) = \sum_{i=1}^{N}
\theta^{\mu\nu}p_i^{\nu}\alpha_i\delta(\tau-\tau_i) \ .
\eeq
The $\cdots$ parts that we do not explicitly write down 
in (\ref{aftkint}) are the contributions involving
the aforementioned extra `contact terms'.  

Our first main combinatorial result shown in 
Appendix B is that: these extra `contact terms' are completely
cancelled when we include the contributions from
$G_2$. This immediately implies that  
\beqa\label{GNC}
\Gamma_N^{\rm ghost}&=& -\left({-ig}\right)^N \sum_{\{\nu_i\}}
\int{dT\over T}\left(\prod_{i=1}^N\int\limits_0^Td\tau_i\right)
e^{{i\over4}\sum_{i<j}p_i\wedge p_j (\nu_i+\nu_j)\eps(\tau_{ij})}
\left(\prod_{j=1}^N (-)^{\alpha_j}\right) \nn\\
&\times& {\cal N}(T)\int{\cal D}x\,
 e^{-\int\limits_0^T{1\over4}{\dot x}^2
(\tau)\,d\tau}\prod_{j=1}^N {\cal O}_j(\tau_j) \ . 
\eeqa

\vspace{0mm}
\begin{minipage}[htb]{15cm}
\begin{center}
\input{vertex.pictex}
\end{center}
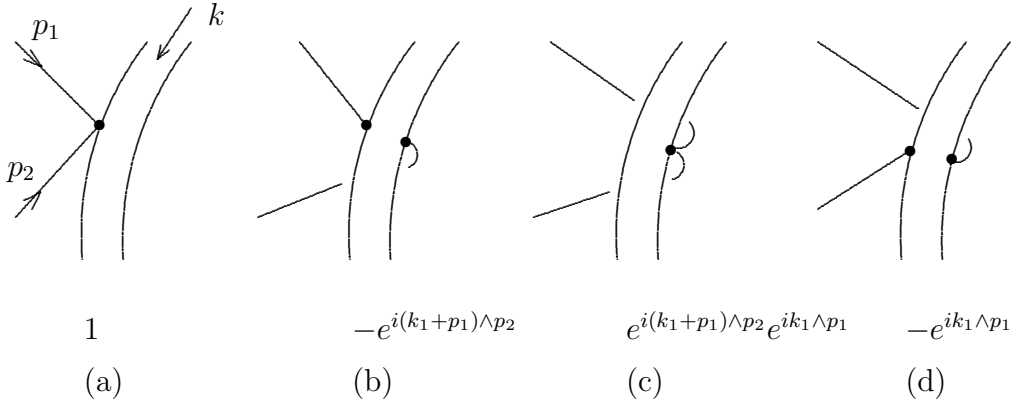
{\bf Figure 1:} {}~Four types of $G_2$ vertices.  
Whenever nonplanar crossings happen, extra
phase factors are attached.   
\end{minipage}
\vspace{8mm}
\subsection{Derivation of Wick contraction rule
for ghost loop}\label{sec3.2}
\indent

We will now derive the following formula:
\beqa\label{Wick}
\Gamma_N^{\rm ghost}&=& -\left({-ig}\right)^N \sum_{\{\nu_i\}}
\int{dT\over T}\left(\prod_{i=1}^N\int\limits_0^Td\tau_i\right)
e^{{i\over4}\sum_{i<j}p_i\wedge p_j (\nu_i+\nu_j)\eps(\tau_{ij})}
\left(\prod_{j=1}^N (-)^{\alpha_j}\right) \nn\\
&\times& 
\Bigl({1\over4\pi T}\Bigr)^{D\over2}
 \Big\langle \prod_{j=1}^N V_j(\tau_j) 
 \Big\rangle_\theta \ , 
\eeqa
where $V_j$'s are the ``gluon''-ghost vertex operators 
\beq\label{phgh}
V_j(\tau)= \epsilon_j^{\rho}\dot x_{\rho}(\tau) 
\exp\Bigl[\, i p_j \cdot x(\tau)\,\Bigr]\ .
\eeq
A rule is that 
$ \langle \prod_{j=1}^N V_j(\tau_j) \rangle_\theta$ 
should be computed with the use of Wick contraction 
in terms of 
\beq
\langle x^\mu(\tau_i) x^\nu(\tau_j) \rangle_\theta = 
- G^{\mu\nu}_{B\theta}(\tau_i,\tau_j;\alpha_i,\alpha_j) ~ ,
\eeq
where the `Green function' is given by \eq{GBtheta}:
\begin{equation}
G^{\mu\nu}_{B\theta}(\tau_i,\tau_j;\alpha_i,\alpha_j) =
   g^{\mu\nu}
G_B(\tau_i,\tau_j)-{i\over T}\theta^{\mu\nu}\alpha_{ij}
  (\tau_i+\tau_j)
-{1\over4T}\alpha^2_{ij}(\theta^2)^{\mu\nu} \ ,
\label{propt}
\end{equation}
$\alpha_{ij} = \alpha_i - \alpha_j$, and
the ordinary worldline Green function
$G_B (\tau_i , \tau_j ) = | \tau_{ij} | - \tau_{ij}^2 / T$. 
This Wick 
contraction rule for the ghost loop is 
remarkably similar to the rules for computing
perturbative string amplitudes in the presence
of a constant NS two-form background field. 
We observe that the propagator (\ref{propt})
fits well with the Seiberg-Witten limit of the
bosonic string worldsheet propagator \cite{kl,u1,u1string}.  
A previous example in commutative space-time that
is close to our analysis can be found in Ref.~\cite{SS} 
for the scalar and spinor QED cases, where the worldline 
formalism is adopted and string-theory-like rules 
for perturbative computations are given.  In the 
noncommutative setup, the $U(1)$ gauge theory behaves
much like nonabelian gauge theories; purely gauge 
degrees of freedom are enough to produce nontrivial 
answers without
adding extra matter fields.  At technical level,
furthermore, to properly take care of the Filk phases, 
interaction vertices should be expanded first before 
performing the path integral over $k (\tau )$.
This forces us to introduce the functional
derivatives in (\ref{GN0C2}) causing extra 
complications here, compared to the commutative 
case analysis.      
   
As an application of the rule, we compute the 
two-point contribution to the amplitude to
obtain: 
\beqa\label{GG2}
\Gamma_2^{\rm ghost}&=&
g^2( \epsilon_1^{\mu}\epsilon_2^{\nu}p_1^{\rho}p_2^{\sigma}
-\epsilon_1^{\mu}p_2^{\nu}\epsilon_2^{\rho}p_1^{\sigma})
\int{dT\over T}\left(1\over 4\pi T \right)^{D\over 2}
\int\limits_0^Td\tau_1\int\limits_0^Td\tau_2 \nn\\
&&\times\Bigl\{
\partial_1G^{\mu\nu}_{B\theta}(\tau_1,\tau_2;\alpha_1,\alpha_2)
\partial_2G^{\rho\sigma}_{B\theta}
(\tau_2,\tau_1;\alpha_2,\alpha_1)       
\Bigr\} \nn\\ 
&&\times\exp\Bigl[\,{1\over2}
\sum_{i.j=1}^2 p_i^\mu G^{\mu\nu}_{B\theta}
(\tau_i,\tau_j;\alpha_i,\alpha_j)\,p_j^\nu\,\,\Bigr]\ ,
\eeqa
where we have used an integration by parts. 
Throughout the rest of this subsection, we will derive
the Wick contraction rule.

For a direct proof of \eq{Wick}, we consider the
evaluation of $x(\tau )$ path integral in \eq{GNC}:
\beq
I\equiv{\cal N}(T)\int\limits_{x(0)=x(T)}\hskip-15pt{\cal D}x 
e^{-\int\limits_0^T{1\over4}\dot{x}^2d\tau}
\prod_{j=1}^N {\cal O}_j(\tau_j) ~ 
\eeq
with the decomposition  
\beq\label{mode}
x^\mu(\tau)=x^\mu_0\, +\, \sum_{n=1}^\infty\, x^\mu_n\,
\sin\left({n\pi\tau\over T}\right)\ ,   
\eeq
satisfying the periodicity condition.
The path integral under this decomposition 
becomes 
\begin{equation}
\int\limits_{x(0)=x(T)}\hskip-15pt{\cal D}x 
 \rightarrow \int_{-\infty}^{\infty} \prod_{n=0}^{\infty}
   dx_n ~ ,
\end{equation}
and we have to evaluate a Gaussian integral for
each $n > 0$.  The zero-mode ($n=0$) integral
gives the momentum conservation condition.
These computations proceed parallel to the 
evaluation of \eq{Xint}, however 
the main difference now is the polynomial (polarization) 
part in ${\cal O}_j(\tau_j)$ involving $\dot{x}$. 
We notice that the Gaussian integrals for modes 
involve a variable shift:
\beq\label{shift}
\dot{x}^\mu(\tau)\ra\dot{x}^\mu(\tau)+{4i\over\pi}
\sum_{k=1}^N p^\mu_k
\sum_{n=1}^\infty{\sin({n\pi\tau_k\over T})
\cos({n\pi\tau\over T})\over n}
-{2\over T}\sum_{k=1}^N\theta^{\mu\nu} p^\nu_k\alpha_k
\sum_{n=1}^\infty\cos({n\pi\tau_k\over T})
\cos({n\pi\tau_k\over T}) ~ ,
\eeq
to `complete the square' in the presence of 
linear terms in $x$ in the exponential coming
from  ${\cal O}_j(\tau_j)$. 
One can show that (\ref{shift}) can be rewritten 
as\footnote{When one evaluates the summation over $n$ in 
the second term of \eq{shift}, there is a subtlety related to
the existence of winding numbers along a circle, which
should be carefully analyzed.  The issue shows up when
one tries to use the formula \eq{formula2} in the
derivation.  In that case, 
one should cut open the loop 
so that $\tau$ becomes larger than all of $\tau_j$ 
to satisfy the condition in \eq{formula2}. 
At the end, in order to join both ends of the
line to form a loop, one should reinstall the  
contribution $\sum_{j=1}^N p_j \eps(\tau-\tau_j)$, which
is zero when $\tau > \tau_j$ for all $j$ due to the
momentum conservation.  At this stage,
one can relax the condition $\tau_j\leq\tau$. 
A safer way is not to use the formula. 
Either way, we can obtain (\ref{amazing}).} 
\beq
\dot{x}^\mu(\tau)\ra 
  \dot{x}^\mu(\tau)-i\sum_{k=1}^{N}p_k^{\nu}\partial_\tau
G^{\mu\nu}_{B\theta}(\tau,\tau_k;\alpha,\alpha_k)
-\Theta^\mu(\tau)\ .
\label{amazing}
\eeq
This shift transforms the polarization part of \eq{OJ}
into  
\beq
\dot{x}^\mu(\tau_j) + \Theta^\mu(\tau_j) 
\ra
\dot{x}_\mu(\tau_j)-i\sum_{k=1(\not=j)}^{N}p_k^{\nu}\partial_j
G^{\mu\nu}_{B\theta}(\tau_j,\tau_k;\alpha_j,\alpha_k)
\equiv\tilde{\cal O}^\mu_j(\tau_j) \ .
\eeq
The second term in $\tilde{\cal O}^\mu_j(\tau_j)$
represents the contractions between $\dot{x}^{\mu}$ and
the exponentials of (\ref{phgh}).  We note that,
even if we deleted $k=j$ (self-contraction) from the 
summation, that contribution actually vanishes.  
We therefore 
arrive at the expression
\beqa\label{ghostN}
\Gamma_N^{\rm ghost}&=& -\left({-ig}\right)^N \sum_{\{\nu_i\}}
\int{dT\over T}\left(\prod_{i=1}^N\int\limits_0^Td\tau_i\right)  
e^{{i\over4}\sum_{i<j}p_i\wedge p_j (\nu_i+\nu_j)\eps(\tau_{ij})}
\left(\prod_{j=1}^N (-)^{\alpha_j}\right) \nn\\
&&\times \,\Bigl({1\over4\pi T}\Bigr)^{D\over2}
\left<\prod_{i=1}^N \epsilon_i\cdot
\tilde{\cal O}_i(\tau_i)\right>
\exp\Bigl[\,{1\over2}\sum_{i.j=1}^N p_i^\mu\, 
 G^{\mu\nu}_{B\theta}
(\tau_i,\tau_j;\alpha_i,\alpha_j)\,p_j^\nu\,\,\Bigr] \ ,
\eeqa
where the contractions between $\dot x$'s in 
$\langle \cdots \rangle$ should be evaluated with 
the ordinary bosonic 
worldline Green function $G_B$. Using the relation
\beq
\langle \dot{x}^\mu(\tau_i)\dot{x}^\nu(\tau_j) \rangle
= \langle \dot{x}^\mu(\tau_i)\dot{x}^\nu(\tau_j)
  \rangle_\theta\ 
\eeq
changes this prescription, and we immediately verify 
that Eq.\eq{ghostN} is exactly the same
as \eq{Wick}. 

\subsection{The gauge loop}\label{sec3.3}
\indent

We analyze the gauge loop in this subsection.  Since the
details are similar to those of the ghost loop case,
we will mainly highlight the genuine features of the
gauge loop.  In addition to $G_1$ and $G_2$, we also have 
the following contributions in the gauge loop case: 
\beqa
G_3(k)&=& 2g \int{d^Dp\over(2\pi)^D} (1-e^{ik\wedge p})\,
p_\alpha \hat{A}_\beta(p) \delta^{\mu\nu}_{\alpha\beta}
\ , \label{G3} \\
G_4(k)&=&2g^2\int{d^Dp_1\over(2\pi)^D}\int{d^Dp_2\over(2\pi)^D}
(1-e^{ik\wedge p_1})(1-e^{i(k+p_1)\wedge p_2})
\hat A_{\alpha}(p_1)\hat A^{\beta}(p_2)
  \delta^{\mu\nu}_{\alpha\beta} \ ,
\label{G4}
\eeqa
which lead to the following path integral expression 
for \eq{gam2}, 
\beq
\Gamma^{\rm gauge} = +{1\over2}\int{dT\over T}
  \int{\cal D}x {\cal D}k\,
{\rm P}\,\exp\Bigl[\,-\int\limits_0^T 
  d\tau \Bigl\{ k^2-ik{\dot x}
+ G_1(k) + G_2(k) +G_3(k)+G_4(k) \Bigr\} \,\Bigr] \ .
\eeq
Here, we introduce $\delta^{\mu\nu}_{\alpha\beta}= 
g^{\mu\alpha}g^{\nu\beta}-g^{\mu\beta}g^{\nu\alpha}$.     
To include both \eq{G3} and the second term of \eq{G1}, 
the functional derivatives in 
\eq{GN0C2} should be modified to 
\beq\label{vertex3}
\epsilon_j^\mu \frac{\delta}
  {\delta(i\dot{x}_\mu(\tau_j))} \quad\ra\quad
\epsilon_j^\alpha \biggl(\,
\I \, \frac{\delta}{\delta(i\dot{x}_\alpha(\tau_j))} \,+\, i\, 
\J_{\alpha\beta} \, p^\beta_j \,\biggr) \ ,
\eeq 
where $\I$ and $\J$ are matrices in the Lorentz space: 
\beq
(\I)^{\mu\nu}=g^{\mu\nu}\ ,\qquad 
(\J_{\alpha\beta})^{\mu\nu} = i\,\delta^{\mu\nu}_{\alpha\beta}\ .
\eeq
After the integration of $k (\tau )$, contributions 
purely from $G_1$ and $G_3$ are expressed as
\beqa\label{PGN0}
\Gamma_N^{(0)\rm gauge}&=&{1\over2}\left({-2g}\right)^N 
  \sum_{\{\nu_i\}}
\int{dT\over T}\left(\prod_{i=1}^N\int\limits_0^T
  d\tau_i\right)
e^{{i\over4}\sum_{i<j}p_i\wedge p_j (\nu_i+\nu_j)
 \eps(\tau_{ij})} 
\left(\prod_{j=1}^N (-)^{\alpha_j}\right) \nn\\
&\times&\mbox{Tr}_L\int{\cal D}x\, e^{i\sum_ip_ix(\tau_i)} 
  \prod_{j=1}^N
\epsilon_j^\alpha \biggl(\,
\I \, \frac{\delta}{\delta(i\dot{x}_\alpha(\tau_j))} +i 
\J_{\alpha\beta} \, p^\beta_j \,\biggr)
\, K\ ,
\eeqa
where $\mbox{Tr}_L$ stands for the Lorentz index trace. 
Similarly, 
the Lorentz index structure of the contact interactions 
$G_2$ plus $G_4$ (see \eq{G2} and \eq{G4}) should also
be generalized 
from $\hat{A}_\mu \hat{A}_\nu g^{\mu\nu}$ to 
\beq\label{vertex4}
\hat{A}_\alpha \hat{A}_\beta 
\Bigl(\, \I \,-\, 2i \, \J_{\mu\nu}\,\Bigr)^{\alpha\beta}\ .
\eeq
Taking the same procedure as the ghost loop case, 
these contact interactions can be encapsulated in the full 
$N$-point expression (Appendix~\ref{ap3}):
\beqa\label{PGN}
\Gamma_N^{\rm gauge}&=& {1\over 2}\left({-ig}\right)^N 
  \sum_{\{\nu_i\}}
\int{dT\over T}\left(\prod_{i=1}^N\int\limits_0^Td\tau_i\right)
e^{{i\over4}\sum_{i<j}p_i\wedge p_j (\nu_i+\nu_j)
\eps(\tau_{ij})}
\left(\prod_{j=1}^N (-)^{\alpha_j}\right) \nn\\
&&\times {\cal N}(T)\int{\cal D}x\,\, 
  e^{-\int\limits_0^T{1\over4}{\dot x}^2
(\tau) d\tau}\mbox{Tr}_L\left(\prod_{j=1}^N \Omat_j(\tau_j) 
  \right)
+\cdots \ ,
\eeqa
where the Lorentz matrix $\Omat_j(\tau)$ is 
\beq\label{OJmat}
\Omat_j(\tau) = \epsilon_j^{\alpha} \biggl(\,
\Bigl\{\,\dot{x}^\alpha(\tau)+\Theta^\alpha(\tau)\,\Bigr\}\,\I
+ 2\J_{\alpha\beta} p^{\beta}_j    \,\biggr) 
\exp\Bigl[\, i p^\rho_j \Bigl(x_\rho(\tau) -
{i\over2}\alpha_j \theta^{\rho\sigma}
  {\dot x}_\sigma(\tau)\Bigr)\,\Bigr]\ .
\eeq
The $\cdots$ parts of (\ref{PGN}), here and from now
on, are just the 
pinching contributions from (\ref{G4}) given by the 
replacement rule
\beq\label{pinch}
(\epsilon_i^\alpha\J_{\alpha\beta}p_i^\beta)
(\epsilon_j^\gamma\J_{\gamma\delta}p_j^\delta)
\quad\ra\quad 
\epsilon_i^\alpha\J_{\alpha\beta}
  \epsilon_j^\beta \ .
\eeq
The proof of this rule is based on
the following combinatorial observation. Let us consider 
the two-point function.  Take two vertices from the
second order term of the exponential series \eq{G3}
and compare with \eq{G4}.  The numerical coefficient 
precisely describes the $G_4$ coupling coefficient: 
${1\over2}(2g)^2 = 2g$.  This matching behavior can be
generalized to arbitrary $N$.  We also note the
following identity, which will be useful in 
the next section as well:
\begin{eqnarray}
& & \int_0^1 \int_0^1 d \tau_i d\tau_j 
e^{{i\over4} (\nu_i + \nu_j ) 
   p_i\wedge p_j\eps(\tau_{ij})}
\delta(\tau_{ij}) \eps (\tau_{ij} ) \nonumber \\ 
& = &  \int_0^1 d \tau_i
\int_0^{\tau_i} d \tau_j e^{ {i\over4}  
 (\nu_i + \nu_j ) 
  p_i\wedge p_j } 
\delta(\tau_{ij})
-   \int_0^1 d \tau_i
\int_{\tau_i}^1 d \tau_j e^{ - {i\over4}  
 (\nu_i + \nu_j ) 
 p_i\wedge p_j } \delta(\tau_{ij}) \nonumber \\  
& = & {1\over2}\int_0^1 d\tau e^{ {i\over4}  
  (\nu_i + \nu_j )  p_i\wedge p_j } 
 - {1\over2}\int_0^1 d\tau  e^{ - {i\over4}  
 (\nu_i + \nu_j )  p_i\wedge p_j } ~ .
\label{xxxx}
\end{eqnarray}
We understand the first formal expression in the 
sense of the point-splitting regularization \cite{sw}.
In (\ref{xxxx}), $\eps( \tau_{ij})$ in the first expression
originates from the antisymmetry under the exchange
of $i$ and $j$ due to the antisymmetric nature of the
matrix $\J_{\alpha \beta}$ (see (\ref{G4}) and
(\ref{pinch})).  Due to the relative $-$ sign
in the last line of ({\ref{xxxx}), a single 
insertion of contact vertex on the inner boundary
($\nu = -1$) comes with a relative $(-)$ sign compared to the
outer insertion ($\nu = +1$) case, precisely the
same as an insertion from $G_3$ (\ref{G3}).   

The rest of the analysis is straightforward. 
After performing the $x$ integration, we formally 
have the $N$-point 
expression 
\beqa
\Gamma_N^{\rm gauge}&=& {1\over2}\left({-ig}\right)^N 
\sum_{\{\nu_i\}}
\int{dT\over T}\left(\prod_{i=1}^N\int\limits_0^Td\tau_i\right)  
e^{{i\over4}\sum_{i<j}p_i\wedge p_j (\nu_i+\nu_j)\eps(\tau_{ij})}
\left(\prod_{j=1}^N (-)^{\alpha_j}\right) \nn\\
&&\times \,\Bigl({1\over4\pi T}\Bigr)^{D\over2}\mbox{Tr}_L
\left<\prod_{i=1}^N \tilde{\Omat}_i(\tau_i)\right>
\exp\Bigl[\,{1\over2}\sum_{i.j=1}^N p_i^\mu\, 
G^{\mu\nu}_{B\theta}
(\tau_i,\tau_j;\alpha_i,\alpha_j)\,p_j^\nu\,\,\Bigr] +\cdots
\eeqa
with the matrix
\beq\label{OJt}
\tilde{\Omat}_j(\tau_j) = \epsilon_j^{\alpha} \biggl(\,
\Bigl\{\,\dot{x}^\alpha(\tau_j)
-i\sum_{k=1(\not=j)}^{N}p_k^{\beta}\partial_j
G^{\alpha\beta}_{B\theta}(\tau_j,\tau_k;\alpha_j,\alpha_k)
\,\Bigr\}\,\I
+ 2\J_{\alpha\beta} p^{\beta}_j  \,\biggr)\ . 
\eeq
Following the same procedures as those in the
ghost loop case, this is summarized as the Wick 
contraction formula
\beqa\label{Wick2}
\Gamma_N^{\rm gauge}&=& {1\over2}(-ig)^N \sum_{\{\nu_i\}}
\int{dT\over T}\left(\prod_{i=1}^N\int\limits_0^Td\tau_i\right)
e^{{i\over4}\sum_{i<j}p_i\wedge p_j (\nu_i+\nu_j)\eps(\tau_{ij})}
\left(\prod_{j=1}^N (-)^{\alpha_j}\right) \nn\\
&\times& \Bigl({1\over4\pi T}\Bigr)^{D\over2}\mbox{Tr}_L
\Big\langle \prod_{j=1}^N \V_j(\tau_j)
 \Big\rangle_\theta +\cdots \ , 
\eeqa
where $\V_j$ is the usual bosonic ``gluon'' vertex operator
\beq\label{gluon}
\V_j(\tau)=\epsilon_j^\alpha\Bigl(\,\dot{x}_\alpha(\tau)\I 
+ 2\J_{\alpha\beta}p^\beta_j\,\Bigr)
\exp\Bigl[\,ip_j\cdot x(\tau)\,\Bigr]\ .
\eeq

As an illustration, we compute the two-point 
contribution from the gauge loop using
(\ref{Wick2}): 
\beqa
\Gamma_2^{\rm gauge}&=&
-{1\over 2}g^2
( \epsilon_1^{\mu}\epsilon_2^{\nu}p_1^{\rho}p_2^{\sigma}
-\epsilon_1^{\mu}p_2^{\nu}\epsilon_2^{\rho}p_1^{\sigma})
\int{dT\over T}\left(1\over 4\pi T \right)^{D\over 2}
\int\limits_0^Td\tau_1\int\limits_0^Td\tau_2 \nn\\
&&\times\Bigl\{
D\partial_1G^{\mu\nu}_{B\theta}(\tau_1,\tau_2;\alpha_1,\alpha_2)
\partial_2G^{\rho\sigma}_{B\theta}
(\tau_2,\tau_1;\alpha_2,\alpha_1)
+8g^{\mu\nu}g^{\rho\sigma}\Bigr\} \nn \\
&&\times\exp\Bigl[\,{1\over2}\sum_{i.j=1}^2 p_i^\mu 
G^{\mu\nu}_{B\theta}
(\tau_i,\tau_j;\alpha_i,\alpha_j)\, p_j^\nu\,\,\Bigr]\ .
\eeqa
Adding this contribution to the ghost contribution
\eq{GG2}, we have the following expression for the
the self-energy part
\beqa
\epsilon_1^\mu\epsilon_2^\nu\Pi_{\mu\nu}&=& -{1\over 2}g^2
( \epsilon_1^\mu\epsilon_2^\nu p_1^{\rho}p_2^{\sigma}
-\epsilon_1^\mu p_2^{\nu}\epsilon_2^\rho p_1^{\sigma})
\int{dT\over T}\left(1\over 4\pi T \right)^{D\over 2}
\int\limits_0^Td\tau_1\int\limits_0^Td\tau_2 \nn\\
&&\times\Bigl\{
(D-2)\partial_1G^{\mu\nu}_{B\theta}
  (\tau_1,\tau_2;\alpha_1,\alpha_2)
\partial_2G^{\rho\sigma}_{B\theta}
(\tau_2,\tau_1;\alpha_2,\alpha_1)
+8g^{\mu\nu}g^{\rho\sigma}\Bigr\} \nn \\
&&\times\exp\Bigl[\,{1\over2}\sum_{i.j=1}^2 p_i^\mu 
G^{\mu\nu}_{B\theta}
(\tau_i,\tau_j;\alpha_i,\alpha_j)\, p_j^\nu\,\,\Bigr]\ .
\label{2point}
\eeqa
This is identical to the results obtained from the
conventional Feynman diagrammatics \cite{ncft,u1} and 
from the Seiberg-Witten limit of the perturbative string
computations \cite{kl,u1,u1string}.  

\section{Emergence of open Wilson lines}\label{sec4}
\setcounter{section}{4}
\setcounter{equation}{0}
\indent

We have developed an efficient method of computing
the multi-point 1PI amplitudes resembling the perturbative
string theory computations.  We now investigate 
how our method helps us obtain the gauge invariant 
completions of various terms. Let us briefly illustrate 
the idea with the two-point function example ( $p_1^{\mu} = - 
p_2^{\mu} = p^{\mu} $ ) \cite{u1}:
\beqa
 \Gamma_2^{NP}&=&  g^2
( \epsilon_1^\mu\epsilon_2^\nu p^{\rho}p^{\sigma}
-\epsilon_1^\mu p^{\nu}\epsilon_2^\rho p^{\sigma})
\int \frac{dT}{T} \left(1\over 4\pi T \right)^{D\over 2}
T^2 \int\limits_0^1 d x \nn\\
&&\times\Bigl\{
- (D-2)  g^{\mu \nu} g^{\rho \sigma} 
(1-2x)^2    +8g^{\mu\nu}g^{\rho\sigma}
 + (D-2) \frac{1}{T^2}
\theta^{\mu\nu} \theta^{\rho\sigma} 
  \Bigr\} \nn \\
&&\times\exp \left( - p^2 T x (1-x) - \frac{1}{4T}
 p \circ p \right) ~ ,
\label{np2p}
\eeqa
written explicitly from (\ref{2point}) for the
nonplanar case with one inner insertion and one
outer insertion.  We will find that our results
immediately produce the gauge invariant completions
of the second term (and their generalizations) 
and the third term in the curly bracket of 
(\ref{np2p}) in the low momentum limit; open Wilson
lines emerge for these types of terms.  
It is not yet clear how to find the 
gauge invariant completion of the first term
even in the low momentum limit.  We will make
further comments on this point later.

The correlation functions from the ghost loop \eq{Wick} 
and the gauge loop \eq{Wick2} can be computed to be: 
\beqa\label{Wick3}
<\prod_{j=1}^N V_j(\tau_j)>_\theta &=&
\exp\Bigl[\, \sum_{i<j}^N \epsilon_i^\mu \epsilon_j^\nu 
{\ddot G}_{B\theta ij}^{\mu\nu}
-i\sum_{i,j=1}^N \epsilon_i^\mu p_j^\nu 
  {\dot G}_{B\theta ij}^{\mu\nu}
\,\Bigr]\biggr|_{m.l.}
e^{{1\over2}\sum_{i,j}p_i {G_{B\theta}}_{ij} p_j} \\
<\prod_{j=1}^N \V_j(\tau_j)>_\theta &=&
\exp\Bigl[\, \Bigl(\sum_{i<j}^N \epsilon_i^\mu 
  \epsilon_j^\nu 
{\ddot G}_{B\theta ij}^{\mu\nu}
-i\sum_{i,j=1}^N \epsilon_i^\mu p_j^\nu 
  {\dot G}_{B\theta ij}^{\mu\nu}\Bigr)\I
+2\sum_{i=1}^N \epsilon _i^\mu \J_{\mu\nu} p_i^\nu
\,\Bigr]\biggr|_{m.l.} \nn \\
&&\times\, e^{{1\over2}\sum_{i,j}p_i {G_{B\theta}}_{ij} p_j} \ .
\label{Wick4}
\eeqa
Here the subscript $m.l.$, i.e., ``multi-linear",
means that we expand the exponential and retain only the
terms which are linear in all $N$-polarization 
vectors \cite{bern}.  The expression $\dot G$ denotes
a derivative with respect to the first argument $\tau$ of
$G$, and $\ddot G$, double derivatives with respect to the
same first argument.
In fact, using more combinatorics,
we can simplify (\ref{Wick4}) further.  Exponentiating 
$\J$ term in \eq{gluon} and taking Wick contractions, 
we can re-express the $\J$ term on the right hand
side of \eq{Wick4} as 
\beq
\exp\Bigl[\, 4 \sum_{i<j}^N 
 (\epsilon_i\J p_i)(\epsilon_j\J p_j) \,\Bigr]
  \biggr|_{m.l.} ~ ,
\eeq
and we apply (\ref{pinch}) to this expression to reproduce 
the pinching contribution parts.  In this way, we can 
absorb the pinching 
part $\cdots$ of (\ref{Wick2}) in a compact form
\beqa
<\prod_{j=1}^N \V_j(\tau_j)>_\theta +\cdots &=&
\exp\Bigl[\, \Bigl(\sum_{i<j}^N \epsilon_i^\mu 
  \epsilon_j^\nu 
{\ddot G}_{B\theta ij}^{\mu\nu} 
-i\sum_{i,j=1}^N \epsilon_i^\mu p_j^\nu 
  {\dot G}_{B\theta ij}^{\mu\nu}\Bigr)\I\nn\\
&&+\,2\sum_{i=1}^N \epsilon_i^\mu \J_{\mu\nu} p_i^\nu 
+4\sum_{i<j}^N \epsilon_i^\mu \J_{\mu\nu} 
  \epsilon_j^\nu\delta(\tau_i-\tau_j)
\,\Bigr]\biggr|_{m.l.} \nn\\
&&\times\, e^{{1\over2}\sum_{i,j}p_i {G_{B\theta}}_{ij} p_j} \ .
\label{gwick}\
\eeqa
The expressions (\ref{Wick3}) and (\ref{gwick}) will
be used for further discussions. 

Let us first consider the polarization dependent part
of (\ref{Wick3}) and (\ref{gwick}).
The $\J$-dependent part of (\ref{gwick}), that
we will call the ($a$) part, generates the second term of 
(\ref{np2p}) at the two-point level.  
When it comes to $\dot G$ and 
$\ddot G$ parts of (\ref{Wick3}) and (\ref{gwick}), there 
are $\theta^{\mu \nu}$-independent terms, which will
be called the ($b$) part.  The ($b$) part is 
responsible for the first term of (\ref{np2p}), upon 
using the integration by parts for $\ddot G$
(see also \cite{HTS}). 
These two parts ($a$) and ($b$) are the
sources for generating the field strength
tensor $\bar{F}_{\mu \nu} 
 = \partial_{\mu} A_{\nu} - \partial_{\nu}A_{\mu}
    + ig [ A_{\mu} , A_{\nu} ]_{\star}$.
In addition, there is an extra contribution from 
$\dot G$
\begin{equation}
 -\frac{1}{T} \sum_{i,j=1}^N \epsilon_i^{\mu} p_j^{\nu}
\theta^{\mu \nu} \alpha_{ij}
=  - \frac{1}{T} \sum_{i,j=1}^N \epsilon_i \wedge p_j
     (\alpha_i - \alpha_j ) 
  = \frac{1}{T} \sum_{i,j=1}^N \epsilon_i \wedge p_j
    \alpha_j  ~ ,
\label{wilso}
\end{equation}
which depends linearly on $\theta^{\mu\nu}$
($\sum_j p_j = 0$).
One should note that, when compared to 
the ($a$) and ($b$) parts,
(\ref{wilso}) comes with a prefactor $1/T$, and it
is responsible for the third term of (\ref{np2p}). 
Similarly, after the scaling $\tau \rightarrow
T \tau$, the term $ \epsilon_i^\mu \J_{\mu\nu} 
  \epsilon_j^\nu\delta(\tau_i-\tau_j)$ of   
(\ref{gwick}) also has the prefactor $1/T$.  
As we increase the number of insertions, each position
moduli integral supplies a factor $T$ after $\tau
\rightarrow T \tau$ scaling.  For the insertions
with the prefactor $1/T$, increasing the number of
insertions does not generate extra powers of $T$, 
and this is an important fact that allows the 
straightforward summation over these terms.  

We now show that for the terms \eq{wilso} ($\theta$-dependent 
part of $\dot{G}$) the higher-point
functions precisely combine to form open Wilson
lines in the low momentum limit.  We will 
concentrate on the gauge loop 
expression for the moment, for the ghost loop 
contribution \eq{Wick3} is a simplified version 
of \eq{gwick}.  Since the external 
insertions are classified into the inner and outer 
boundary insertions, we introduce a momentum flow 
between two boundaries  
\beq
  k= \sum_{r=1}^{N_1} p_r =  - \sum_{a=1}^{N_2} p_a 
 =   - \sum_{i=1}^N \alpha_i p_i 
\eeq
where $N_1$ and $N_2$; $N=N_1+N_2$ are the number of 
insertions on outer ($N_1$, $\alpha = 0$, 
$\nu = 1$; $r,s, \cdots$) and inner 
($N_2$, $\alpha = 1$, $\nu = -1$; $a,b, \cdots$) 
boundaries, respectively.  
Scaling the moduli parameters $\tau_i\ra T\tau_i$ and 
using an identity following from the momentum
conservation, 
\beq
- i \sum_{i,j=1}^N p_i \wedge p_j ~\alpha_{ij}
(\tau_i+\tau_j)
= - i \sum_{i,j=1}^N p_i \wedge p_j ~(\nu_i+\nu_j)~\tau_{ij}~~ .
\eeq
we derive the following formula
for the $N$-point amplitudes:
\beqa\label{GNbulk}
\Gamma_{N,\{\nu_i\}}^{\rm gauge} &=& {1\over 2}
\int{dT\over T}\left({1\over4\pi T}\right)^{D\over2} 
  \exp \left[ -m^2 T - \frac{k \circ k}{4T} \right] \nn \\
&&\times   
(-igT)^{N1}\left( \prod_{r=1}^{N_1}\int\limits_0^1 d\tau_r\right)
\exp \left( + {i\over2}\sum_{r<s }p_r\wedge p_s \eps(\tau_{rs})
  -  i p_r \wedge p_s \tau_{rs} \right) \nn\\
&&\times
(igT)^{N2}\left(\prod_{a=1}^{N_2}\int\limits_0^1 d\tau_a\right) 
\exp \left( - {i\over2}\sum_{a<b}p_a\wedge p_b \eps(\tau_{ab})
   + i p_a \wedge p_b \tau_{ab} \right)   \nn\\
&&\times
\mbox{Tr}_L\,\exp\Bigl[\, \Bigl(\sum_{i<j}^N 
  \epsilon_i^\mu \epsilon_j^\nu 
{\ddot G}_{B\theta ij}^{\mu\nu} 
-i\sum_{i,j=1}^N \epsilon_i^\mu 
  p_j^\nu {\dot G}_{B\theta ij}^{\mu\nu}\Bigr)\I\nn\\
&&+\,2\sum_{i=1}^N \epsilon_i^\mu \J_{\mu\nu} p_i^\nu 
+ \frac{4}{T} \sum_{i<j}^N \epsilon_i^\mu \J_{\mu\nu} 
  \epsilon_j^\nu\delta(\tau_i-\tau_j)
\,\Bigr]\biggr|_{m.l.}  
 \exp ( \sum_{i<j} p_i \cdot p_j G_B )\ .
\eeqa
In this expression, we have introduced the IR cutoff mass
$m^2$.  If one wants to use our $U(1)$ gauge
theory to simulate the broken $U(1)$ by separating 
two (bosonic) D-branes, the mass $m$ could be 
interpreted as being proportional to the separation
distance.  As assumed at the outset, we neglect the 
$\theta^{\mu \nu}$-independent
$\I$ parts in (\ref{GNbulk}), since they are the 
($b$) part terms.  Furthermore, by 
taking the low momentum
limit, the $G_B = T ( | \tau_{ij} | - \tau_{ij}^2 ) $ part 
will also be neglected.  

We first sum up the terms which have zero number of
$\J$-insertions.  From the 
$\theta^{\mu \nu}$-dependent $\I$ part, we 
derive 
\beq
\exp\Bigl[ - \,{1\over T}\Bigl(\sum_{r=1}^{N_1} 
\epsilon_r \wedge k
+\sum_{a=1}^{N_2} \epsilon_a\wedge 
  k \Bigr)\I\,\Bigr]\biggr|_{m.l.}
= (-1)^{N_1 + N_2} T^{-N_1-N_2} 
    \left( \prod_{r=1}^{N_1} 
  \epsilon_r\wedge k\right)
\left(\prod_{a=1}^{N_2} \epsilon_a\wedge k\right) \I \ .
\eeq
It is important to note that it comes with negative
powers of $T$, which cancels the positive powers of
$T$ coming from the position moduli integrals.  We also
note that the corresponding ghost contribution has 
the same expression as this except for the Lorentz unit 
matrix $\I$.  Since the two sets of position moduli 
integrations, the inner and the outer, 
describe the phase parts of the
generalized $\star$-products, 
$\star_{N_1}$ and $\star_{N_2}$ \cite{trek2,klp,mehen};
\beq
J_N(p_1,...,p_N;k)=\left(\prod_{i=1}^N\int_0^1 d\tau_i\right)
\exp \left(  {i\over2}\sum_{a<b}^N p_a\wedge p_b 
 \{\eps(\tau_{ab})
   - 2 \tau_{ab}\} \right)\ ,
\label{stark}
\eeq
and the effective action sums up with the 
following combinatorics
\beq
\Gamma[A]=\sum_{N=0}^{\infty}{1\over N!}
\sum_{\{\nu_i\}}\Gamma_{N,\{\nu_i\}} 
=\sum_{N_1=0}^\infty\sum_{N_2=0}^\infty{1\over N_1! N_2!}
\Gamma_{N,\{\nu_i\}}\ ,
\eeq
the effective action can be expressed in terms of the straight 
open Wilson line \cite{trek2}:
\beqa\label{OWL}
W_k[A] &=& \sum_{N=0}^\infty{(ig)^N\over N!}
\int{d^Dp_1\over(2\pi)^D}\cdots\int{d^Dp_N\over(2\pi)^D}
(2\pi)^D\delta^D(k-\sum_{i=1}^N p_i) \nn\\
&\times&\Bigl[\,J_N(p_1,\cdots,p_N;k)\,
(l\cdot \hat{A})(p_1) \cdots (l\cdot\hat{A})(p_N)\,\Bigr] \ ,
\eeqa
where 
\beq
l^\mu = \theta^{\mu\nu} k_\nu \ ,
\eeq
living on each boundary.  
The planar contributions $N_1=0$ or $N$ vanish because 
$l=0$ (no momentum flow between two boundaries).
The completely factorized $T$-integral provides
the `closed string 
propagator' between two boundaries ($n=0$ for all
$N$): 
\beq\label{prop}
{\cal K}_n (k) := 
\int_0^\infty{dT\over T}
\left({1\over4\pi T}\right)^{D\over2}T^n             
\exp \left[\,-m^2 T - \frac{k \circ k}{4T} \,\right] \ ,
\eeq
which can be straightforwardly evaluated to yield
an expression involving modified Bessel functions 
$K_n (z)$
\begin{equation}
{\cal K}_n (k) = 2 \left( \frac{1} {4 \pi} 
  \right)^{\frac{D}{2}} \left( \frac{1}{2}
\sqrt{ \frac{k \circ k }{ m^2} } \right)^{n - \frac{D}{2}}
 K_{n - \frac{D}{2} } ( \sqrt{ m^2 k \circ k } ) ~ . 
\end{equation}
In the above, we have also defined its derivatives 
(arbitrary positive $n$) for
later convenience.   Combining the calculations
so far, 
we obtain the effective action in the following form:
\beq
\Gamma[A]={D-2\over 2}
\int{d^D k\over(2\pi)^D}W_k[A]\,{\cal K}_0(k)\, W_{-k}[A] ~~,
\eeq
where we have included the ghost contribution.  
The outer boundary ($N_1$-summation) produces the
factor $W_k [A]$, while the inner boundary
gives the factor $W_{-k} [A]$ ($N_2$-summation),
which is Hermitian conjugate to $W_k [A]$. 

Next, we turn our attention to the terms 
containing nonzero number of insertions from
the $\J$ parts in \eq{GNbulk}, which are also of our interest. 
The ghost part does not contribute to this case. 
As noticed in Appendix~\ref{ap3}, 
the $\delta (\tau_{ij} )$-contact term exists only for pairs 
inserted on the 
same boundary (figures (a) and (c) in {}~Figure 1); 
for other types of insertions, they cancel out due to 
$\nu_i + \nu_j = 0$ in (\ref{xxxx}). 
Hence these parts can also be factorized as 
\beq\label{Jterm}
\exp \left( 2\sum_{r=1}^{N_1} \epsilon_r^\mu \J_{\mu\nu} p_r^\nu 
+ \frac{4}{T} \sum_{r<s}^{N_1} \epsilon_r^\mu \J_{\mu\nu} 
  \epsilon_s^\nu\delta(\tau_{rs}) \right) \ ,
\eeq
and a similar form for the other boundary. 
The first summation in \eq{Jterm} is identical to 
$\der_\nu A_\mu - \der_\mu A_\nu$ with a plane wave substitution 
$A^\mu\ra\sum_{r=1}^{N_1}\epsilon^\mu_r \exp ( ip_r\cdot x )$. 
On the same ground, the second summation corresponds 
to a commutator form 
accompanied by the Filk phase factor 
$\exp[ {i\over2}\sum_{r<s}p_r\wedge p_s\eps(\tau_{rs})]$, 
which yields the $\star$-commutator between
two $A$'s in a contact term (see (\ref{xxxx})). 
When there are $N = 2 n_{\rm contact} + n_{\rm cubic}$
insertions along one boundary, except for 
the Filk phase factor for the $\star$-commutator, 
the would-be $\star_{N}$-kernel
$J_N$ of (\ref{stark}) rearranges itself to 
$\star_{N^{\prime}}$-kernel 
$J_{N^{\prime}}$, where
$N^\prime =  n_{\rm contact} + n_{\rm cubic}$, because of 
the $\delta(\tau_{rs})$ part of a contact term insertion.  
In other words,
even if the $\star$-commutator part involves two
insertions of $A$'s, it counts as a single insertion
when it comes
to the $\star_N$-kernel.  The same is true for the
counting of the power of $T$ due to the extra
$1/T$ factor for the contact term.  After all we notice 
that \eq{Jterm} is nothing but the Fourier transform 
of $2\bar{F}_{\mu\nu}$ as naturally expected from \eq{Sgauge}. 
The descendents of an open Wilson line are thus defined 
as follows  
\beqa
\F_k^{(n)}[A]&=& 2^n (ig)^n \sum_{N=0}^\infty
  {(ig)^N\over N!}\left(\prod_{i=1}^{n+N}
\int{d^D p_i\over(2\pi)^D}\right)\hat{\matF}
 (p_1)\cdots\hat{\matF}(p_n)
(l\cdot\hat{A})(p_{n+1})\cdots(l\cdot\hat{A})(p_{n+N}) \nn\\
&&\times J_{n+N}(p_1,\cdots,p_{n+N};k) 
(2\pi)^D\delta^D(k-\sum_{i=1}^{n+N}p_i) \ ,
\eeqa
where $(\matF)_{\mu\nu}={\bar F}_{\mu\nu}$ containing
the $\star$-commmutator term, and obviously 
$\F_k^{(0)}[A]=W_k[A]$. Dividing $N$ into 
$ ( {\tilde N}_1 + n_1 ) + ( {\tilde N}_2 + n_2  ) $, 
where $n_i$; ($i=1,2$) represent the numbers of 
outer/inner $\matF$ insertions, 
the combinatorics for the effective action reads 
\beq
\Gamma [A] = 
\sum_{N=0}^\infty{1\over N!}\sum_{\{\nu_i\}}\Gamma_{N,\{\nu_i\}}
=  \sum_{\tilde{N}_1}\sum_{\tilde{N}_2}
\sum_{n_1}\sum_{n_2}{1\over \tilde{N}_1! \tilde{N}_2! n_1! n_2!}
\Gamma_{N,\{\nu_i\}}\ .
\eeq 
These considerations immediately yield
\beq\label{Gfermi}
\Gamma [A]= {1\over2}{\rm{Tr}}_L 
 \sum_{n_1=0}^\infty \sum_{n_2=0}^\infty
{(-1)^{n_1} \over n_1! n_2!}\int {d^D k\over(2\pi)^D}
\F_{k}^{(n_1)}[A]~{\cal K}_{n_1+n_2}(k)~ \F_{-k}^{(n_2)}[A] \ .
\eeq
The result is precisely the gauge invariant completion
of field strength tensors in terms of the insertion of
an open Wilson line for each boundary.

An outstanding issue is whether one can find 
the simple gauge-invariant completion of the terms 
involving the field strength coming from the ($b$) part.
There are two sources of complications for the
computations; first, the terms $\dot G \cdots \dot G$
appear to perturb the expressions for the $\star_N$
kernel.  Secondly, the integration by parts 
involved in turning $\ddot G$'s into $\dot G$'s,
in general, can generate extra terms.  
Under any circumstances, it remains
to be seen if the analog of the ($b$) part will be 
present in the supersymmetric setup.   
In fact, the terms computed in \cite{trek1} (and further
considered in \cite{trek2}) appear to be 
related to the terms from the ($a$) part involving the
four $\J$'s.  It is amusing to note that the 
``gluon" vertex operator of (\ref{gluon}) 
is formally similar to the 0-picture gauge
boson vertex operator 
\begin{equation}
{\cal V}^{ 0} = g_o (2 \alpha^{\prime} )^{-1/2}
 t^a ( i \dot X^{\mu} + 2 \alpha^{\prime}
 k_\nu \psi^{\nu} \psi^{\mu} ) e^{i k \cdot X}
\end{equation}
of superstring theory, where $t^a$ is the 
Chan-Paton matrix and $g_o$ is the open
string coupling, {\em once} we replace the Fermion
bilinear $\psi^{\nu} \psi^{\nu}$ with 
$\J^{\mu \nu}$.  The terms considered in \cite{trek1}
actually originate from the $\psi^{\nu} \psi^{\mu}$
part of the 0-picture vertex operators. Regarding
the $\J$ matrix as the bilinear of worldline fermion 
fields in our formalism produces the precisely the 
same answer as that of \cite{trek1}.  It will
also be interesting to understand this connection
closely, for example, by constructing the 
supersymmetric version of our formulation.

%
%
%
%

\par
\section*{Acknowledgements}
We would like to thank Prof. Soo-Jong Rey for inspiring
discussions.  We are also grateful to Chong-Oh Lee
for the participation at the early stage of this work. 
Y. Kiem was supported by the KRF Grant 2001-015-DP0082, 
Y. Kim by the BK21 project of the Ministry of Education, 
and H.-T. S. by the KOSEF Brain Pool Program.

\newpage
\appendix
\section*{Appendix}
\section{The stripping method}\label{ap1}
\setcounter{equation}{0}
\indent

In this appendix, we present the details of the stripping
method.  For simplicity, we choose to consider the 
noncommutative real scalar field theory with a cubic
interaction as a concrete example:
\beq
S=\int d^Dx\Bigl( {1\over2}(\der\phi)^2 +{1\over2}m^2\phi^2 + 
{g\over3!}\phi\star\phi\star\phi \,\Bigr)(x)\ .
\eeq
Decomposing ${\hat\phi}$ (Fourier transform of $\phi$) into
classical ${\hat\phi}_0$ and quantum ${\hat\varphi}$ fields and
adopting the procedure of \cite{Roiban}, 
we have the following one-loop relevant part:
\beqa\label{p3act}
S^{1-loop}&=&\int{d^Dk_1\over(2\pi)^D}{d^Dk_2\over(2\pi)^D}
  (2\pi)^D
\Bigl\{\, {1\over2}(k_1^2+m^2)\delta^D(k_1+k_2) \nn\\
&+&{g\over4}\int{d^Dp\over(2\pi)^D}\delta^D(k_1 + k_2+p)
\,(e^{{i\over2}k_1\wedge p}+e^{-{i\over2}k_1\wedge p})
{\hat\phi}_0(p)
\,\Bigr\}{\hat\varphi}(k_1){\hat\varphi}(k_2)\ .
\eeqa
If one regards $\exp[\pm{i\over2}k_1\wedge p]\hat{\phi}_0$ 
terms as planar and nonplanar interactions, one has to 
precisely go through the computation of \cite{ana1}.
In this case, there is no phase factor stripping 
process (the nonstripping method). 

The stripping method is based on the following phase 
space observation. 
From the formulae for general bosonic functions  
\beqa
\int \varphi_1\star\phi_0\star\varphi_2(x)\,d^Dx 
&=&{1\over(2\pi)^{2D}}
\int d^Dk_1 d^D k_2 d^Dp \delta^D(k_1+k_2+p)\nn\\
&\times& e^{-{i\over2}k_1\wedge p}\,
{\hat \varphi_1}(k_1){\hat \phi_0}(p)
{\hat \varphi_2}(k_2)\ ,\label{base1}\\
\int \varphi_1 \star \phi_1 \star \phi_2 \star\varphi_2(x)\,d^Dx 
&=&{1\over(2\pi)^{3D}} \int d^Dk_1 d^D k_2 d^Dp_1 d^Dp_2 
\delta^D(k_1+k_2+p_1+p_2) \nn \\
&\times& e^{-{i\over2}(k_1\wedge p_1 + p_2\wedge k_2) }\,
{\hat \varphi_1}(k_1){\hat \phi_1}(p_1)
{\hat \phi_2}(p_2){\hat \varphi_2}(k_2)\ ,
\label{base2}
\eeqa
appropriate {}~Fourier bases for the noncommutative 
determinant appear to be $e^{-{i\over2}k_1\wedge p}$ 
for a cubic vertex (this might be interpreted as a phase 
factor for 
functional derivatives of second order) and to be 
$e^{-{i\over2}(k_1\wedge p_1 + p_2\wedge k_2) }$ for a 
contact vertex if exists. 
We hence remove this factor from \eq{p3act} in the 
three-body interaction. 
This should be understood as an inclusion of $\star$-operation 
into the background field (denoted as 
${\hat\phi_{0\star}}(p)$); we thus have 
\beq\label{strip}
\Gamma = \ln\mbox{Det}^{-{1\over2}}\Bigl[\, 
(k_i^2 +m^2)\delta^D(k_i+k_j) +
{g\over2}\int{d^Dp\over(2\pi)^D}\,\delta^D
(k_i+k_j+p)\,(1+e^{i k \wedge p})
{\hat\phi_{0\ast}}(p) \,\Bigr]\ .
\eeq
Alternatively we have an option of not 
including $e^{-{i\over2}k_1\wedge p}$ 
into the background field (nonstripping method):
\beq\label{nostrip}
\Gamma = \ln\mbox{Det}^{-{1\over2}}
\Bigl[\, (k_i^2 +m^2)\delta^D(k_i+k_j) +
{g\over2}\int{d^Dp\over(2\pi)^D}\,\delta^D(k_i+k_j+p)\,
(e^{{i\over2} k \wedge p}+e^{-{i\over2} k \wedge p})
{\hat\phi_0}(p) \,\Bigr]\ .
\eeq
This time, the products between the background fields should
be understood as conventional commuting products.  
This approach was examined 
in~\cite{ana1}, and we will not repeat it here.

Further computations following the approach
based on \eq{strip} should be in order. 
Even if this process is almost parallel to the nonstripping 
method, a subtlety should be taken care of; namely,
one should first define the notion of $\star$-products
for background fields in the presence of two boundaries
in the double-line notation.  At tree level, the number
of boundary is one.  We expand the action assuming a 
path ordered exponential: 
\beqa\label{p3G}
\Gamma &=& {1\over2}\int{dT\over T}\int{\cal D}x{\cal D}k
\exp\Bigl[\,-
\int\limits_0^T(k^2+m^2-ik{\dot x})d\tau\,\Bigr]\nn\\
&\times&\sum_{N=0}^\infty\sum_{n=0}^N\Bigl(-{g\over2}\Bigr)^N\, 
\prod_{l=1}^n\int\limits_0^{\tau_{l+1}}d\tau_l
\int{d^Dp_l \over(2\pi)^D}\,{\hat\phi}_{0\star}(p_l)
\prod_{j=1}^{N-n}\int\limits_0^{\tau'_{j+1}}d\tau'_j
\int{d^Dp'_j\over(2\pi)^D}\,
{\hat\phi}_{0\star}(p'_j) \nn\\
&\times&\exp\Bigl[\,-i\sum_{j=1}^{N-n}p'_j 
\wedge k(\tau'_j)\,\Bigr] \ ,
\eeqa
where $\tau_{n+1}$ and $\tau'_{N+1-n}$ are equal to $T$ 
for given $n$ and $N$. 
The $p_l$ and $p'_j$ are the external momenta 
corresponding to vertex insertions in either outer or inner 
boundaries.  We shall assign the sign factor $\nu_l=1$ 
to the outer insertion case 
and $\nu_j=-1$ to the inner insertion case. 
The {}~Feynman amplitudes are defined 
as functions of the set of external momenta 
\beqa
\{\,p_i\,\}&=& \{\,p_l \quad\mbox{for}\quad i=1,2,
\cdots,n\ ;\quad   
p'_j \quad\mbox{for}\quad i=n+1,\cdots,N \,\}\ ,\label{mset}\\
\{\,\tau_i\,\}&=& \{\,\tau_l \quad\mbox{for}
\quad i=1,2,\cdots,n\ ;\quad   
\tau'_j \quad\mbox{for}\quad i=n+1,\cdots,N \,\}\ . 
\label{tset}
\eeqa
Corresponding to the product 
$\prod_{i=1}^N{\hat\phi_{0\star}}(p_i)$ in \eq{p3G}, 
it is necessary to replace the whole product with 
\beq
\phi_0(x(\tau_1))\star^\nu
\phi_0(x(\tau_2))\star^\nu\cdots\star^\nu\phi_0(x(\tau_N))
\eeq
in the configuration space when performing the 
plane wave substitution 
$\phi_0 \ra \sum_{n=1}^N \exp[ip_n x]$, 
where $\nu=1$ is applied to the products from 
$\phi_0(x(\tau_1))$ 
to $\phi_0(x(\tau_n))$, $\nu=-1$ to those from 
$\phi_0(x(\tau_{n+1}))$ 
to $\phi_0(x(\tau_N))$, and $\nu=0$ to the products between 
those two sets.  According to the Chan-Paton charge
assignment in string theory, for example, when a charge 
is attached to one boundary, an anti-charge should be 
attached to the other boundary.
Remembering that 
$N!$ overcountings occur by definition in the plane 
wave substitution, 
we obtain for $N$ point function part as 
\beqa\label{p3rule}
&&\prod_{l=1}^n\int\limits_0^{\tau_{l+1}}
d\tau_l{\hat\phi}_{0\ast}(p_l)
\prod_{j=1}^n\int\limits_0^{\tau'_{j+1}}d\tau'_j
{\hat\phi}_{0\star}(p'_j) \nn\\
&&\ra\,{1\over N!}\,
\prod_{l=1}^n\int\limits_0^{\tau_{l+1}}d\tau_l
\prod_{j=1}^{N-n}\int\limits_0^{\tau'_{j+1}}d\tau'_j
\Bigl(\, e^{ip_1x(\tau_1)}\star^\nu
e^{ip_2x(\tau_2)}\star^\nu\cdots
\star^\nu e^{ip_Nx(\tau_N)} \nn \\
&&+ \quad(\mbox{all $p_i$ permutations})\, \Bigr)\ . 
\eeqa
Furthermore, since we want to take account of all
orderings of $\tau$ 
as $\star$-product effects, it is very natural to 
incorporate $\tau$ dependence 
into $\star^\nu$-product by defining 
\beqa\label{newstar}
e^{ip_ix(\tau_i)}\star^\nu e^{ip_jx(\tau_j)}&\define&
\exp\Bigl[\,-{i\over2}\nu\eps(\tau_{ij})\theta^{\mu\nu}
\der^\mu_y\der^\nu_z\,\Bigr]\,
e^{ip_iy}e^{ip_jz}\Bigr|_{y=x(\tau_i),z=x(\tau_j)} \nn\\
&=& \exp\Bigl[\,{i\over2}p_i\wedge p_j \nu\eps(\tau_{ij})\,\Bigr]
e^{ip_ix(\tau_i)+ip_jx(\tau_j)}\ ,
\eeqa
where $\nu$ is related to the average 
\beq\label{nui+j}
\nu={\nu_i + \nu_j \over2} \ ,
\eeq
and
\beq\label{tij}
\tau_{ij}= \tau_i - \tau_j \ 
\eeq
(see \cite{sw} as well for a related string theory
discussion). The symbol $\eps (x)$ picks up the sign
of its argument $x$ and it will be typically understood
via the point-splitting regularization.
Interchanging integration 
variables $\tau_i$, all permutation terms 
in \eq{p3rule} can be arranged as all possible ordered 
integrals having 
the same integrand. We thus conclude that the 
right hand side of \eq{p3rule} is  
\beq
{1\over N!}\,
\sum_{\{\nu_i\}}\int\limits_0^Td\tau_N\cdots
\int\limits_0^Td\tau_1\,
\Xi\,\prod_{j=1}^N e^{ip_jx(\tau_j)}\ ,
\eeq
where the summation on $\{\nu_i\}$ denotes the sum
over all possible 
outer/inner insertions.  Here, the `stripped' Filk
phase is given by  
\beq\label{Xi}
\Xi = \exp\Bigl[\,{i\over4}\sum_{i<j}^Np_i
\wedge p_j(\nu_i+\nu_j)
\eps(\tau_{ij})\,\Bigr] \ .
\eeq
The $N$-point amplitudes ($\Gamma\ft\sum{1\over N!}\Gamma_N$)
are therefore obtained as 
\beqa\label{p3GN}
\Gamma_N&=&{1\over2}\left({-g\over2}\right)^N \sum_{\{\nu_i\}}
\int{dT\over T}
\left(\prod_{i=1}^N\int\limits_0^Td\tau_i\right)\,\Xi
\nn\\
&&\times\int{\cal D}x e^{i\sum_ip_ix(\tau_i)}
\int{\cal D}k 
 e^{-\int\limits_0^T(k^2(\tau)+m^2-ik{\dot x}(\tau))d\tau}
\prod_{j=1}^N e^{-i\alpha_j p_j\wedge k(\tau_j)}\ ,
\eeqa
where $\alpha_j$ takes either 0 or 1 for outer and inner 
boundary insertions, respectively;
\beq\label{aj}
\alpha_j = {1 - \nu_j \over2} \ .
\eeq
Note that in this expression we have new quantities 
$\Xi$ and $\alpha_j$ instead of $\nu_{j}$, which do 
not appear in the nonstripping method~\cite{ana1}. 
After performing the $k$ integration, we compute the 
remaining $x$ integration (the counterpart in the 
nonstripping method is written as $X$): 
\beq\label{Xint}
{\tilde X}\equiv\int\limits_{x(0)=x(T)}\hskip-15pt{\cal D}x\,
e^{-{1\over4}\int\limits_0^T {\dot x}^2 d\tau}
\prod_{j=1}^N \exp\Bigl[\, i{p_j}^\mu \Bigl(x_\mu(\tau_j) - 
{i\over2}\alpha_j\theta_{\mu\nu}
{\dot x}^\nu(\tau_j) \Bigr)\,\Bigr]\ .
\eeq
In the present case, we have to use three of the following 
formulae (assuming $0< \tau_i \pm \tau_j <2 T$):
\beqa
&&\sum_{n=1}^\infty{\cos nx \over n^2} =
{1\over4}(|x|-\pi)^2-{\pi^2\over12}\ ,
\qquad 0\leq x \leq 2\pi \label{formula1}\\
&&\sum_{n=1}^\infty{\sin nx \over n} ={1\over2}(\pi-x)\ ,
\qquad 0\leq x < 2\pi \label{formula2}\\
&&\sum_{n=1}^\infty\cos n(x-a)  =\pi\delta(x-a)-{1\over2}\ .
\qquad a-\pi<x<a+\pi \label{formula3}
\eeqa
In contrast, we use only two of them in the nonstripping 
method; we use the derivative of \eq{formula1} 
instead of \eq{formula2} (See Eq.(2.11) in Ref.~\cite{ana1}). 
Note that the second formula is the only source of
$\wedge$-product phase factors.  Direct computation
yields
\beqa
{\tilde X}&=&\left({1\over4\pi T}\right)^{D\over2}\exp\Bigl[\,\,
-{T\over4}\sum_{i,j=1}^N p_i\cdot p_j\,
\Bigl\{\,\Bigl(\,1-{|\tau_i-\tau_j|\over T}\,\Bigr)^2
-\Bigl(\,1-{\tau_i+\tau_j\over T}\,\Bigr)^2\,\Bigr\}\nn\\
&-&{i\over2}\sum_{i,j=1}^N p_i\wedge p_j\,
\Bigl\{\,\alpha_j\Bigl(\,1-{\tau_i-\tau_j\over T}\,\Bigr)
+\alpha_j\Bigl(\,1-{\tau_i+\tau_j\over T}\,\Bigr)\,\Bigr\}\nn\\
&-&{1\over4T}\sum_{i,j=1}^N p_i\circ 
p_j\,\alpha_i\alpha_j\,\,\Bigr]\ ,
\eeqa
or more compactly
\beq
{\tilde X}=
\left({1\over4\pi T}\right)^{D\over2}
\exp\Bigl[\,{1\over2}\sum_{i.j=1}^N G^{\mu\nu}_{B\theta}
(\tau_i,\tau_j;\alpha_i,\alpha_j)\,p_i^\mu p_j^\nu\,\,\Bigr]\ ,
\eeq
where $G_{B\theta}$ is the noncommutative version of $G_B$:
\beq\label{GBtheta}
G^{\mu\nu}_{B\theta}(\tau_i,\tau_j;\alpha_i,\alpha_j)=g^{\mu\nu}
G_B(\tau_i,\tau_j)-{i\over T}
\theta^{\mu\nu}\alpha_{ij}(\tau_i+\tau_j)
-{1\over4T}\alpha^2_{ij}(\theta^2)^{\mu\nu}\ ,
\eeq
and $\alpha_{ij}=\alpha_i-\alpha_j$. 
By comparing the result with \cite{ana1}, we establish 
the relation between two methods  
\beq
X = \Xi \, {\tilde X} \ ,
\eeq
and we verify that \eq{p3GN} leads to the same results as 
those presented in~\cite{ana1}. 
The equivalence between these two methods was verified
only at one-loop level.  To tackle multi-loops using 
nonstripping method, more subtle `branch choice' similar
to the one given by the Appendix of \cite{trek1} appears
necessary.  The stripping method is safer in this sense,
since the overall Filk phase ($\star^{\nu}$ with
multiple boundaries) can be unambiguously
determined within the 
purely field theory considerations \cite{ana2,Roiban}.

\section{Cancellation of $G_2$ in ghost loop}\label{ap2}
\setcounter{equation}{0}
\indent

As seen in \eq{GN0C}, the nonplanar phase contributions 
from $G_1$ are 
summarized by the phase factor
$e^{-i\alpha_j p_j\wedge k(\tau_j)}$ at each 
vertex position $\tau_j$. 
When two vertices converge into one position 
\beq\label{contact}
e^{-i\alpha_1 p_1 \wedge k(\tau_1)}\, e^{-i\alpha_2 p_2 
 \wedge k(\tau_2)}
\qquad \ra \qquad
e^{-i\alpha_1 p_1 \wedge k(\tau_1)}\, 
e^{-i\alpha_2 p_2 \wedge (k(\tau_1)+ p_1) }\ ,
\eeq
where the momentum conservation should be taken 
into account along
with the point-splitting regularization 
(note that $k(\tau_1)\not=k(\tau_2)$ as $\tau_1 \rightarrow
\tau_2$ when $\tau_1 > \tau_2$). 
This identification \eq{contact} 
holds independently of whether two vertices are 
converging to a point or remain separated, as long as
there are not any extra insertions between them. 
In view of this relation, the phase factors in $G_2$ are 
in fact the same as those in $G_1^2$ form. 
This is clearly seen in the original form 
of $S_2$ in \eq{S2}, and is rather trivial in the nonstripping 
approach. Hence we simply attach the overall phase factor 
$\Xi$, where 
a pair of converging positions $\tau_i$ and $\tau_j$ are to be 
understood as having an infinitesimal separation so that 
$\eps(\tau_{ij})$ can reproduce the phase factors for 
intertwining pairs. 
In this way, the phase part of $(G_1)^2$ and $G_2$ perfectly 
match when two points converge into one vertex. 
We only have to discuss the rest; namely, we can 
concentrate on the cancellation problem of  
$\delta$-functions produced by functional derivatives,
simply omitting phase factors by hand. 

The above argument leads us to examine the (commutative) 
scalar QED case, which is known as a fact that all 
four point contributions are contained in the 
$\delta$-function of a worldline two-point function: 
$\dot{x}^\mu(\tau_1)\dot{x}^\nu(\tau_2) = 
2g^{\mu\nu}\delta(\tau_{12}) + 
\langle \dot{x}^\mu(\tau_1)\dot{x}^\nu(\tau_2)
\rangle_{regular}$. 
Our purpose is to show how the $\delta$-functions 
(extra `contact terms') generated by functional derivative 
operation in \eq{GN0C2} are canceled. 
The full action that we consider in view of the
previous argument is given by 
\beq\label{scaG}
\Gamma^{\rm scalar} = +\int{dT\over T}\int{\cal D}x {\cal D}k\,
{\rm P}\,
\exp\Bigl[\,-\int\limits_0^T d\tau \Bigl\{ k^2-ik{\dot x}
+ G_1(k) + G_2(k) \Bigr\} \,\Bigr] \ ,
\eeq
with
\beqa
G_1(k)&=& k^2+2gk^{\mu}
\int{d^D p\over(2\pi)^D} \hat A_{\mu}(p)\ , \\
G_2(k)&=&g^2\int{d^D p_1 \over(2\pi)^D}\int{d^D p_2\over(2\pi)^D}
\hat A_{\mu}(p_1) \hat A_{\nu}(p_2)g^{\mu\nu} \ .
\eeqa
Instead of \eq{GN0C2} and \eq{K}, let us consider the 
following expressions:
\beq\label{scaG0}
\Gamma_N^{(0)\rm} = \left({-2g}\right)^N 
\int{dT\over T}\left(\prod_{i=1}^N\int\limits_0^Td\tau_i\right) 
\int{\cal D}x\, e^{i\sum_ip_ix(\tau_i)}  
\prod_{j=1}^N \epsilon_j^{\mu}\,
\frac{\delta}{\delta (i\dot x_{\mu}(\tau_j))}\, \K\ ,
\eeq
and
\beq\label{sK}
\K = {\cal N}(T)\, e^{-{1\over4}
\int\limits_0^T{\dot x}^2 d\tau} \ .
\eeq
We can also write down a formula for the $N$-point 
contributions originated 
from purely $G_2$ parts (i.e. no inclusion of three-point 
vertex contributions): 
\beqa\label{scaG1}
\Gamma^{(1)}&=&\sum_{n=0}^\infty{(-g^2)^n\over (2n)!}
\int{dT\over T}
\int{\cal D}k{\cal D}x\, e^{-\int(k^2-ik\dot{x})}  \nn\\
&\times&
\prod_{i,j=1,i\not=j}^n \int\limits_0^Td\tau_i d\tau_j 
\int{d^Dp_i d^Dq_j\over (2\pi)^{2D}}
\hat{A}_\mu(p_i)\hat{A}_\nu(q_j)
g^{\mu\nu}\delta(\tau_i-\tau_j)\ ,
\eeqa
where we have used the fact that the number of ways of 
inserting $\delta$-functions 
$(2n-1)!!$ times the number of shuffling
external legs $(2n)!!$ is equal to $N!$ ($N=2n$). 
We thus find: 
\beq\label{scaG1N}
\Gamma^{(1)}_N = (ig)^N
\int{dT\over T}\int{\cal D}x\,e^{i\sum_ip_ix(\tau_i)} 
\left(\prod_{i=1}^N \int\limits_0^T d\tau_i \right) 
{2^n \over n!} \Bigl(\sum_{i<j}^N
\epsilon_i\cdot\epsilon_j
\delta(\tau_i-\tau_j)\Bigr)^n\Bigr|_{m.l.}\K\ ,
\eeq
where the subscript $m.l.$ means that only the 
contributions ``multi-linear" in all $N$ polarization 
vectors should be retained.
In the $N=2$ case, the $\delta$-function contribution 
from \eq{scaG0}  
\beq\label{N2G0}
(-2g)^N\frac{\delta^2\,\K}
{\delta(i\dot{x}^\nu(\tau_2))\delta(i\dot{x}^\mu(\tau_1))}
=g^2\Bigl(\,2g^{\mu\nu} \delta(\tau_1-\tau_2) 
- \dot{x}^\mu(\tau_1)\dot{x}^\nu(\tau_2)\,\Bigr)\,\K
\eeq
cancels the \eq{scaG1N}. Thus, 
$\Gamma_2=\Gamma^{(0)}_2+\Gamma^{(1)}_2$ is given by 
$\langle \dot{x}^\mu(\tau_1)\dot{x}^\nu(\tau_2) \rangle$.

It is useful to see $N=3$ case, where we can observe 
the cross term cancellation between $G_1$ and $G_2$. 
Though there is not any $\Gamma^{(1)}_3$ contribution in this 
case, 
we have $\Gamma^{(1)}_2$ and $\Gamma^{(0)}_1$ combination 
(denote $\Gamma^{(0+1)}_N$), whose integrand is given by 
multiplying those two integrands. From 
$\Gamma^{(0)}_3$, we have 
\beq\label{N3G0}
\frac{(-2g)^N  \delta^3\, \K}
{\delta(i\dot{x}^\rho_3)
\delta(i\dot{x}^\nu_2)\delta(i\dot{x}^\mu_1)}
=-ig^3\Bigl(\,  2\delta_{12}\dot{x}^\rho_3 g^{\mu\nu}
       + 2\delta_{13}\dot{x}^\nu_2 g^{\mu\rho}
       + 2\delta_{23}\dot{x}^\mu_1 g^{\rho\nu} 
+ \dot{x}^\mu_1\dot{x}^\nu_2\dot{x}^\rho_3 \,\Bigr)\,\K\ ,
\eeq
where $\dot{x}^\mu_i\equiv\dot{x}^\mu(\tau_i)$ and 
$\delta_{ij}\equiv\delta(\tau_{ij})$. From the cross 
terms, we have the integrand for $\Gamma^{(0+1)}_3$:
\beq
\biggl((-2g){i\over2}\dot{x}^\mu_1\biggr)\biggl((ig)^2 2
\epsilon_2 \cdot\epsilon_3\delta_{23}\biggr)
+\mbox{cyclic permutations} \ .
\eeq
Again the $\delta$-function 
cancellation happens in 
$\Gamma_3=\Gamma^{(0)}_3+\Gamma^{(0+1)}_3$, 
which is thus given by 
$\langle \dot{x}^\mu_1\dot{x}^\nu_2\dot{x}^\rho_3 \rangle$.
Note also that \eq{N2G0} and \eq{N3G0} are regular 
parts themselves. 

In the $N=4$ case, we have three kinds of contributions 
$\Gamma^{(0)}_4$, 
$\Gamma^{(1)}_4$ and $\Gamma^{(0+1)}_4$, whose integrand is 
given by those of 
$\Gamma^{(0)}_2$ and $\Gamma^{(1)}_2$: 
\beqa\label{N4G0}
\Gamma^{(0)}_4 &\sim&
(-2g)^4\epsilon_1^\mu\epsilon_2^\nu\epsilon_3^\rho
\epsilon_4^\sigma\frac{\delta^4 \,\K}
{\delta (i\dot{x}_1^\mu)\delta (i\dot{x}_2^\nu)\delta 
(i\dot{x}_3^\rho)\delta (i\dot{x}_4^\sigma)}\nn\\
&=& 2g^4
\Bigl(\sum_{i<j}\epsilon_i\cdot\epsilon_j
\delta(\tau_i-\tau_j)\Bigr)^2 \biggr|_{m.l.}
    -2g^4 \Bigl(\sum_{i<j}
(\epsilon_i\cdot\epsilon_j)\delta_{ij}\Bigr)
\Bigl( \sum_{k<l} \epsilon_k\cdot\dot{x}_k 
\epsilon_l\cdot\dot{x}_l\Bigr)\biggr|_{m.l.}
\nn\\
&+&g^4\epsilon_1\cdot\dot{x}_1\epsilon_2\cdot\dot{x}_2
\epsilon_3\cdot\dot{x}_3\epsilon_4\cdot\dot{x}_4\ ,
\eeqa
\beqa\label{N4G01}
\Gamma^{(0+1)}_4 &\sim& \biggl(g^2\sum_{i<j}^4
(\epsilon_i^{\mu_i} \epsilon_j^{\mu_j})
(2g^{\mu_i\mu_j}\delta_{ij} 
- \dot{x}^{\mu_i}_i\dot{x}^{\mu_j}_j)\biggr)
\biggl((ig)^2\sum_{j<k}^4 
2\epsilon_j\cdot\epsilon_k 
\delta_{jk}\biggr)\biggr|_{m.l.} \nn\\
&=&-4g^4\Bigl(\sum_{i<j}^4 \epsilon_i \cdot \epsilon_j 
\delta_{ij} \Bigr)^2\biggr|_{m.l.} 
 +2g^4 \Bigl(\sum_{i<j} \epsilon_i\cdot\epsilon_j 
\delta_{ij}\Bigr)
\Bigl( \sum_{k<l} \epsilon_k\cdot\dot{x}_k 
 \epsilon_l\cdot\dot{x}_l\Bigr)\biggr|_{m.l.}\ ,
\eeqa
\beqa\label{N4G1}
\Gamma^{(1)}_4 &\sim& (ig)^4 2 
\Bigl(\sum_{i<j}^4 
\epsilon_i\cdot\epsilon_j \delta_{ij} \Bigr)^2\biggr|_{m.l.} \\
&=&4g^4\Bigl((\epsilon_1\cdot\epsilon_2)
(\epsilon_3\cdot\epsilon_4)\delta_{12}\delta_{34}
+(\epsilon_1\cdot\epsilon_3)
 (\epsilon_2\cdot\epsilon_4)\delta_{13}\delta_{24}
+(\epsilon_2\cdot\epsilon_3)
 (\epsilon_1\cdot\epsilon_4)
 \delta_{23}\delta_{14}\,\Bigr)\ .\nn
\eeqa 
Gathering these up, all $\delta$-functions cancel except 
for the part
$\langle \dot{x}_1^\mu\dot{x}_2^\nu\dot{x}_3^\rho
 \dot{x}_4^\sigma \rangle$, which gives 
$\Gamma_4$. It is not difficult to generalize the 
above expressions for higher values of 
$N$; 
one can check the cancellations in the cases of
higher values of $N$ straightforwardly. 

Finally, in order to confirm the $N$-point function \eq{GNC}, we 
have to replace \eq{sK} with \eq{K} in the above argument. 
The derivative of $K$ then generates an 
additional $\Theta^\mu(\tau)$ term 
(see \eq{defth} for definition): 
\beq\label{dK}
\frac{\delta K}{\delta(i\dot{x}_\mu(\tau_1))}
={i\over2}\Bigl(\, \dot{x}^\mu(\tau_1) + 
\Theta^\mu(\tau_1)\,\Bigr )\,K
\define {i\over2}v^\mu(\tau_1)\,K\ .
\eeq
One may wonder if the derivative singularities proportional 
to $\Theta^\mu$ 
survive. However $\Theta^\mu$ appears symmetric in the 
exchange $\dot{x}^\mu(\tau)
\lra \Theta^\mu(\tau)$ as understood from \eq{dK}. Since 
all derivative 
singularities proportional to $\dot{x}$ vanish in the above 
argument, this kind of new terms are also canceled 
out in a parallel way.  We therefore conclude that the 
total $N$-point contribution of a ghost loop 
is given by \eq{GNC}. 

\section{Cancellation of $G_2$ in gauge loop}\label{ap3}
\setcounter{equation}{0}
\indent

Denoting $G_2+G_4$ contribution as $\Gamma^{(1)}$ similar 
to \eq{scaG1}, let us define 
\beqa\label{qG1}
\Gamma^{\rm(1) gauge}&=&
{1\over2}\sum_{n=0}^\infty{(-g^2)^n\over (2n)!}
\int{dT\over T}
\int{\cal D}k{\cal D}x\, e^{-\int(k^2-ik\dot{x})}\,
\prod_{i,j=1,i\not=j}^n \int\limits_0^Td\tau_i d\tau_j \nn\\
&\times& \mbox{Tr}_L\,
\int{d^Dp_i d^Dq_j\over (2\pi)^{2D}}
\hat{A}_\alpha(p_i)\hat{A}_\beta(q_j)
(\, \I \,-\, 2i \, \J_{\mu\nu}\,)^{\alpha\beta}
\delta(\tau_i-\tau_j) \ .
\eeqa
The changes from the ghost loop case (Appendix B) are 
the overall factor $-\,{1\over2}$ and the Lorentz 
structure \eq{vertex4}. Here the local phase factors of 
four Feynman diagram combinations in Figure.~1 are ignored 
for the same reason as before, and 
the overall phase factor $\Xi$ is also omitted for 
simplicity: one should notice that this omission is 
irrelevant if one follows the nonstripping approach 
of Appendix~\ref{ap1}. The $N (=2n)$ point functions 
for \eq{qG1} are then defined as 
\beqa
\Gamma^{\rm(1) gauge}_N &=& {1\over2}(ig)^N
\int{dT\over T}\int{\cal D}x\,e^{i\sum_ip_ix(\tau_i)} 
\left(\prod_{i=1}^N \int\limits_0^T d\tau_i \right) \nn\\
&\times& {1\over n!}\mbox{Tr}_L\,
\Bigl(\sum_{i,j}^N\eps^\alpha_i \eps_j^\beta 
(\I-2i\J_{\mu\nu})^{\alpha\beta}
\delta(\tau_i-\tau_j)\Bigr)^n\Bigr|_{m.l.}\K\ . \label{naiveG1}
\eeqa
Suppose that a pair of converging points are separated by point 
splitting, and the overall phase factor $\Xi$ is then attached 
in the above expression. In the case when a pair of external 
lines are inserted on different boundaries, the pair does 
not contribute to $\Xi$ (see (\ref{xxxx}) with
$\nu_i + \nu_j = 0$); such kinds of four-point contribution 
from \eq{naiveG1} vanish due to the antisymmetric nature 
of the $\J$ matrix. 
Other four-point contributions from the $\J$ term survive. 
These nonvanishing contributions have nothing to do with 
the following argument on cancellations of the
derivative-induced $\delta$-functions.  We hence 
consider the following expression for this purpose:
\beq\label{qG1N}
\Gamma^{\rm(1) gauge}_N = {1\over2}(ig)^N
\int{dT\over T}\int{\cal D}x\,e^{i\sum_ip_ix(\tau_i)} 
\left(\prod_{i=1}^N \int\limits_0^T d\tau_i \right) 
{D2^n\over n!}
\Bigl(\sum_{i<j}^N\epsilon_i \cdot \epsilon_j
\delta(\tau_i-\tau_j)\Bigr)^n\Bigr|_{m.l.}\K + \cdots \ .
\eeq
Let us consider the $\Gamma^{(0)}$ part without phase factors,
\beqa\label{qG0N}
\Gamma_N^{(0)\rm gauge}&=&{1\over2}
\left({-2g}\right)^N \sum_{\{\nu_i\}}
\int{dT\over T}
\left(\prod_{i=1}^N\int\limits_0^Td\tau_i\right) \nn\\
&\times&\mbox{Tr}_L\int{\cal D}x\, 
e^{i\sum_ip_ix(\tau_i)} \prod_{j=1}^N
\epsilon_j^\alpha \biggl(\,
\I \, \frac{\delta}{\delta(i\dot{x}_\alpha(\tau_j))} +i 
\J_{\alpha\beta} \, p^\beta_j \,\biggr)
\, \K\ ,
\eeqa
and define the cross term contribution 
$\Gamma^{\rm(0+1) gauge}_N$ 
whose integrand is given by multiplying those of 
$\Gamma^{\rm(0) gauge}_{N-n}$ and $\Gamma^{\rm(1) gauge}_n$:  
\beqa\label{qG01N}
\Gamma^{\rm(0+1) gauge}_N &=& {1\over2}
\sum_{n=1}^{[{N\over2}]}(-2g)^{N-n}(ig)^n{2^n\over n!}
\int{dT\over T}\int{\cal D}x\,e^{i\sum_ip_ix(\tau_i)} 
\left(\prod_{i=1}^N \int\limits_0^T d\tau_i \right) \nn\\
&\times&\mbox{Tr}_L\prod_{l=1}^{N-n} \epsilon_l^\alpha 
\biggl(\,\I \, \frac{\delta}{\delta(i\dot{x}_\alpha(\tau_l))} +i 
\J_{\alpha\beta} \, p^\beta_l \,\biggr)\, 
\Bigl(\sum_{i<j}^{2n}\epsilon_i \cdot \epsilon_j \I
\delta(\tau_i-\tau_j)\Bigr)^n\biggr|_{m.l.} \K \ ,
\eeqa
where the orderings of contact term insertions should be 
understood properly, although it is not explicit here. 
In the same way as in Appendix B, the 
$\Gamma^{\rm(1) gauge}_N$ and 
$\Gamma^{\rm(0+1) gauge}_N$ contributions cancel the 
$\delta$-functions produced by 
$\frac{\delta}{\delta i\dot{x}}$ 
in \eq{qG0N}.  It is straightforward to explicitly verify
this up to $N=4$, and further generalizations
are also possible.   
Gathering \eq{qG1N}, \eq{qG0N} and \eq{qG01N}, 
we therefore conclude that the $N$-point function is 
given by \eq{PGN}. 


\end{document}